\documentclass[twocolumn]{aastex63}
\def \eg {e.g.}

\def \ie {i.e.}
\def \cf {cf.}
\def \lcdm {{\hbox{$\Lambda$CDM}}}
\def \omegam {{\hbox{$\Omega_m$}}}
\def \omegal {{\hbox{$\Omega_\Lambda$}}}
\def \hzero {{\hbox{$H_0$}}}
\def \arcmin {\hbox{$^\prime$}}
\def \arcsec {\hbox{$^{\prime\prime}$}}
\def \deg {\hbox{$^\circ$}}
\def \nh {\hbox{$N_{\rm H}$}}

\def \msun {\hbox{${\rm M_\odot}$}}

\def \mfive {\hbox{$M_{500}$}}
\def \lfive {\hbox{$L_{500}$}}
\def \rfive {\hbox{$r_{500}$}}

\newcommand{\ergs }{\mbox{erg s$^{-1}$}}

\newcommand{\kmsmpc }{\mbox{km s$^{-1}$ Mpc$^{-1}$}}

\newcommand{\kev }{\mbox{keV}}

\newcommand{\mujyb }{\mbox{$\mu$Jy beam$^{-1}$}}

\newcommand{\whz }{\mbox{W Hz$^{-1}$}}
\newcommand{\acisi }{ACIS-I}

\newcommand{\acis }{ACIS}

\newcommand{\obsid }{ObsID}

\newcommand{\uv }{\textit{uv}}

\newcommand{\wsclean }{\textsc{WSClean}}

\newcommand{\killms }{\textsc{killMS}}
\newcommand{\ddfacet }{\textsc{DDFacet}}

\newcommand{\xspec }{\textsc{xspec}}

\newcommand{\ciao }{\textsc{ciao}}

\newcommand{\caldb }{\textsc{caldb}}

\newcommand{\lira }{LIRA}

\newcommand{\chandra }{{\em Chandra}}

\newcommand{\rosat }{ROSAT}

\newcommand{\ugmrt }{uGMRT}

\newcommand{\vla }{VLA}
\newcommand{\vlaE }{Very Large Array}
\newcommand{\jvla }{JVLA}

\newcommand{\lofar }{LOFAR}

\newcommand{\wsrt }{WSRT}
\newcommand{\wsrtE }{Westerbork Synthesis Radio Telescope}

\newcommand{\lotss }{LoTSS}
\newcommand{\lotssE }{LOFAR Two-meter Sky Survey}

\newcommand{\tgss }{TGSS}
\newcommand{\tgssE }{TIFR GMRT Sky Survey}

\newcommand{\sdss }{SDSS}

\usepackage{graphicx}
\usepackage{amssymb}
\usepackage{multirow,bigdelim}
\usepackage{txfonts}
\shorttitle{The beautiful mess in Abell 2255}
\shortauthors{Botteon et al.}

\begin{document}

\title{The beautiful mess in Abell 2255}

\correspondingauthor{Andrea Botteon}
\email{botteon@strw.leidenuniv.nl}

\author[0000-0002-9325-1567]{A. Botteon}
\affiliation{Leiden Observatory, Leiden University, PO Box 9513, 2300 RA Leiden, The Netherlands}
\affiliation{INAF - IRA, via P.~Gobetti 101, 40129 Bologna, Italy}

\author{G. Brunetti}
\affiliation{INAF - IRA, via P.~Gobetti 101, 40129 Bologna, Italy}

\author{R. J. van Weeren}
\affiliation{Leiden Observatory, Leiden University, PO Box 9513, 2300 RA Leiden, The Netherlands}

\author{T. W. Shimwell}
\affiliation{ASTRON, the Netherlands Institute for Radio Astronomy, Postbus 2, 7990 AA Dwingeloo, The Netherlands}
\affiliation{Leiden Observatory, Leiden University, PO Box 9513, 2300 RA Leiden, The Netherlands}

\author{R. F. Pizzo}
\affiliation{ASTRON, the Netherlands Institute for Radio Astronomy, Postbus 2, 7990 AA Dwingeloo, The Netherlands}

\author{R. Cassano}
\affiliation{INAF - IRA, via P.~Gobetti 101, 40129 Bologna, Italy}

\author{M. Iacobelli}
\affiliation{ASTRON, the Netherlands Institute for Radio Astronomy, Postbus 2, 7990 AA Dwingeloo, The Netherlands}

\author{F. Gastaldello}
\affiliation{INAF - IASF Milano, via A.~Corti 12, 20133 Milano, Italy}

\author{L. B\^{\i}rzan}
\affiliation{University of Hamburg, Hamburger Sternwarte, Gojenbergsweg 112, 21029 Hamburg, Germany}

\author{A. Bonafede}
\affiliation{Dipartimento di Fisica e Astronomia, Universit\`{a} di Bologna, via P.~Gobetti 93/2, 40129 Bologna, Italy}
\affiliation{INAF - IRA, via P.~Gobetti 101, 40129 Bologna, Italy}

\author{M. Br\"{u}ggen}
\affiliation{University of Hamburg, Hamburger Sternwarte, Gojenbergsweg 112, 21029 Hamburg, Germany}

\author{V. Cuciti}
\affiliation{University of Hamburg, Hamburger Sternwarte, Gojenbergsweg 112, 21029 Hamburg, Germany}

\author{D. Dallacasa}
\affiliation{Dipartimento di Fisica e Astronomia, Universit\`{a} di Bologna, via P.~Gobetti 93/2, 40129 Bologna, Italy}
\affiliation{INAF - IRA, via P.~Gobetti 101, 40129 Bologna, Italy}

\author{F. de Gasperin}
\affiliation{University of Hamburg, Hamburger Sternwarte, Gojenbergsweg 112, 21029 Hamburg, Germany}

\author{G. Di Gennaro}
\affiliation{Leiden Observatory, Leiden University, PO Box 9513, 2300 RA Leiden, The Netherlands}

\author{A. Drabent}
\affiliation{Th\"{u}ringer Landessternwarte, Sternwarte 5, 07778 Tautenburg, Germany}

\author{M. J. Hardcastle}
\affiliation{ Centre for Astrophysics Research, University of Hertfordshire, College Lane, Hatfield AL10 9AB, UK}

\author{M. Hoeft}
\affiliation{Th\"{u}ringer Landessternwarte, Sternwarte 5, 07778 Tautenburg, Germany}

\author{S. Mandal}
\affiliation{Leiden Observatory, Leiden University, PO Box 9513, 2300 RA Leiden, The Netherlands}

\author{H. J. A. R\"{o}ttgering}
\affiliation{Leiden Observatory, Leiden University, PO Box 9513, 2300 RA Leiden, The Netherlands}

\author{A. Simionescu}
\affiliation{SRON Netherlands Institute for Space Research, Sorbonnelaan 2, 3584 CA Utrecht, The Netherlands}
\affiliation{Leiden Observatory, Leiden University, PO Box 9513, 2300 RA Leiden, The Netherlands}
\affiliation{Kavli Institute for the Physics and Mathematics of the Universe (WPI), The University of Tokyo, Kashiwa, Chiba 277-8583, Japan}



\begin{abstract}
\noindent
We present \lofar\ observations of one of the most spectacular objects in the radio sky: Abell 2255. This is a nearby ($z = 0.0806$) merging galaxy cluster hosting one of the first radio halos ever detected in the intra-cluster medium (ICM). The deep \lofar\ images at 144 MHz of the central $\sim10$ Mpc$^2$ region show a plethora of emission on different scales, from tens of kpc to above Mpc sizes. In this work, we focus on the innermost region of the cluster. Among the numerous interesting features observed, we discover remarkable bright and filamentary structures embedded in the radio halo. We incorporate archival \wsrt\ 1.2 GHz data to study the spectral properties of the diffuse synchrotron emission and find a very complex spectral index distribution in the halo spanning a wide range of values. We combine the radio data with \chandra\ observations to investigate the connection between the thermal and non-thermal components by quantitatively comparing the radio and X-ray surface brightness and the spectral index of the radio emission with the thermodynamical quantities of the ICM. Despite the multitude of structures observed in the radio halo, we find that the X-ray and radio emission are overall well correlated. The fact that the steepest spectrum emission is located in the cluster center and traces regions with high entropy possibly suggests the presence of seed particles injected by radio galaxies that are spread in the ICM by turbulence generating the extended radio halo.
\end{abstract}

\keywords{radiation mechanisms: non-thermal -- radiation mechanisms: thermal -- galaxies: clusters: individual (Abell 2255) -- galaxies: clusters: general -- galaxies: clusters: intracluster medium}

\section{Introduction}

Mergers between galaxy clusters are very energetic events capable of releasing up to $\sim10^{64}$ ergs on a few Gyr timescale in the intra-cluster medium (ICM). A significant fraction of the energy involved in cluster mergers is dissipated by weak shocks and turbulence, heating the gas \citep[\eg][]{markevitch07rev}, while a smaller percentage can be channelled into non-thermal components, namely relativistic particles and magnetic fields, that reveal themselves through Mpc-scale diffuse radio emission \citep[\eg][for a review]{vanweeren19rev}. The presence of diffuse synchrotron sources associated with the ICM raises fundamental questions about the processes responsible for the conversion of the energy dissipated by large-scale motions generated during cluster mergers into particle acceleration mechanisms on much smaller scales \citep[\eg][for a review]{brunetti14rev}. According to the current leading scenario, turbulence is believed to be responsible for the formation of the central and roughly spherical sources called \textit{radio halos}, while shocks play a role in the generation of the elongated and polarized sources found in cluster outskirts known as \textit{radio relics}. \\
\indent
The galaxy cluster Abell 2255 (hereafter A2255) is a nearby ($z=0.0806$) system in a complex dynamical state. The merger dynamics is still unclear from optical observations \citep[\eg][]{burns95, yuan03, golovich19atlas}, and possibly involves multiple substructures colliding along different merger axes. Multiple X-ray observations show an ICM emission slightly elongated in the E-W direction and the presence of temperature asymmetries, confirming the unrelaxed state of the system \citep[\eg][]{feretti97, davis98, sakelliou06, akamatsu17a2255}. A2255 was one of the earliest clusters to be found to host a diffuse radio halo at its center \citep{jaffe79}. Detailed studies of the radio halo and relic (located in the NE) have been performed with the \wsrtE\ \citep[\wsrt;][]{feretti97, pizzo08, pizzo09, pizzo11} and with the \vlaE\ \citep[\vla;][]{govoni05, govoni06}, unveiling complex structure in the diffuse radio emission, which extends up to large distances from the center of the galaxy cluster. In addition, a large number of tailed radio galaxies are found in the cluster environment, among which, some are embedded in the halo emission \citep[\eg][]{harris80eight, feretti97, miller03a2255}. A summary of the properties of A2255 collected from the literature is reported in Tab.~\ref{tab:properties}.

\begin{table}[t]
 \centering
 \caption{Properties of A2255 derived from the literature: redshift \citep[$z$;][]{struble99}, equatorial coordinates, mass within \rfive\ \citep[\mfive;][]{planck16xxvii}, luminosity in the $0.5-2.0$ \kev\ band within \rfive\ \citep[\lfive;][]{eckert17xcop}, central entropy \citep[$K_0$;][]{cavagnolo09}, virial temperature \citep[$kT_{vir}$;][]{hudson10}, radio halo power at 1.4 GHz \citep[$P_{1.4}$;][]{govoni05}.}
 \label{tab:properties}
  \begin{tabular}{lr} 
  \hline
  \hline
  $z$ & 0.0806 \\
  Right ascension (h, m, s) & 17 12 31 \\
  Declination (\deg, \arcmin, \arcsec) & $+$64 05 33  \\
  \mfive\ ($10^{14}$ \msun) & $5.38\pm0.06$  \\
  \lfive\ ($10^{44}$ \ergs) & $2.08\pm0.02$ \\
  $K_0$ (kev cm$^2$) & $529\pm28$ \\
  $kT_{vir}$ (keV) & $5.8\pm0.2$ \\
  $P_{1.4}$ ($10^{23}$ \whz) & $9.0\pm0.5$ \\
  \hline
  \end{tabular}
\end{table}

In this paper, we report on the results from new \lofar\ High Band Antennas (144 MHz) observations of A2255 together with archival \wsrt\ (1.2 GHz) and \chandra\ data. This manuscript is organized as follows. In Section~\ref{sec:observations}, we give an overview of the observations and data reduction. In Section~\ref{sec:results}, we present the radio and X-ray images and the results from the surface brightness and spectral analysis. In Section~\ref{sec:discussion}, we discuss our findings and compare them with other literature works. In Section~\ref{sec:conclusions} we summarize and present our conclusions. Throughout the paper, we adopt a \lcdm\ cosmology with $\omegal = 0.7$, $\omegam = 0.3$ and $\hzero = 70$ \kmsmpc, in which 1 arcsec corresponds to 1.512 kpc at the cluster redshift, and use the convention $S_\nu \propto \nu^{-\alpha}$ for radio synchrotron spectrum.

\section{Observations and data reduction}\label{sec:observations}

\subsection{\lofar}

A2255 was observed by \lofar\ as part of the \lotssE\ \citep[\lotss;][]{shimwell17, shimwell19}, an ongoing deep low-frequency survey of the Northern sky. Each \lotss\ pointing consists of 8~hr observations in the $120-168$ MHz frequency band, centered at 144 MHz. The field of A2255 was covered by three \lotss\ pointings (P254+65, P258+63, P260+65), which were processed with the data reduction pipeline\footnote{\url{https://github.com/mhardcastle/ddf-pipeline}} v2.2 developed by the \lofar\ Surveys Key Science Project \citep[see Section 5 in][]{shimwell19}. The pipeline carries out direction-dependent calibration with \killms\ \citep{tasse14arx, tasse14, smirnov15} and performs imaging of the entire \lofar\ field-of-view (FoV) with \ddfacet\ \citep{tasse18}. \\
\indent
We further enhanced the quality of the images of the field covering A2255 by performing additional phase and amplitude self-calibration loops in a square region of 1 deg$^2$ centered on the target to mitigate for residual artifacts left over by the automated pipeline. All sources outside this region were subtracted out from the \uv-data. A self calibration cycle was then performed on the combined dataset after correction for the \lofar\ station beam at the target position. This extraction and re-calibration scheme has been already applied in recent \lofar\ works \citep[\eg][]{hardcastle19, cassano19, botteon19lyra}, and it will be discussed in more detail in a forthcoming paper (van Weeren et al., in prep.). \\
\indent
Final imaging was performed with \wsclean\ v2.6 \citep{offringa14} by using suitable taperings of the visibilities and Briggs weightings \citep{briggs95} to obtain images at different resolutions. We checked the \lofar\ flux density scale by comparing the brightest compact sources extracted from the \tgssE-Alternative Data Release \citep[\tgss-ADR1;][]{intema17} with the \lofar\ image. No flux density scale offset was found. We adopt a conservative calibration error of 20\% on \lofar\ flux density measurements, as was done by \lotss\ \citep{shimwell19}. 

\begin{figure*}[t]
 \centering
 \includegraphics[width=\hsize,trim={0cm 0cm 0cm 0cm},clip]{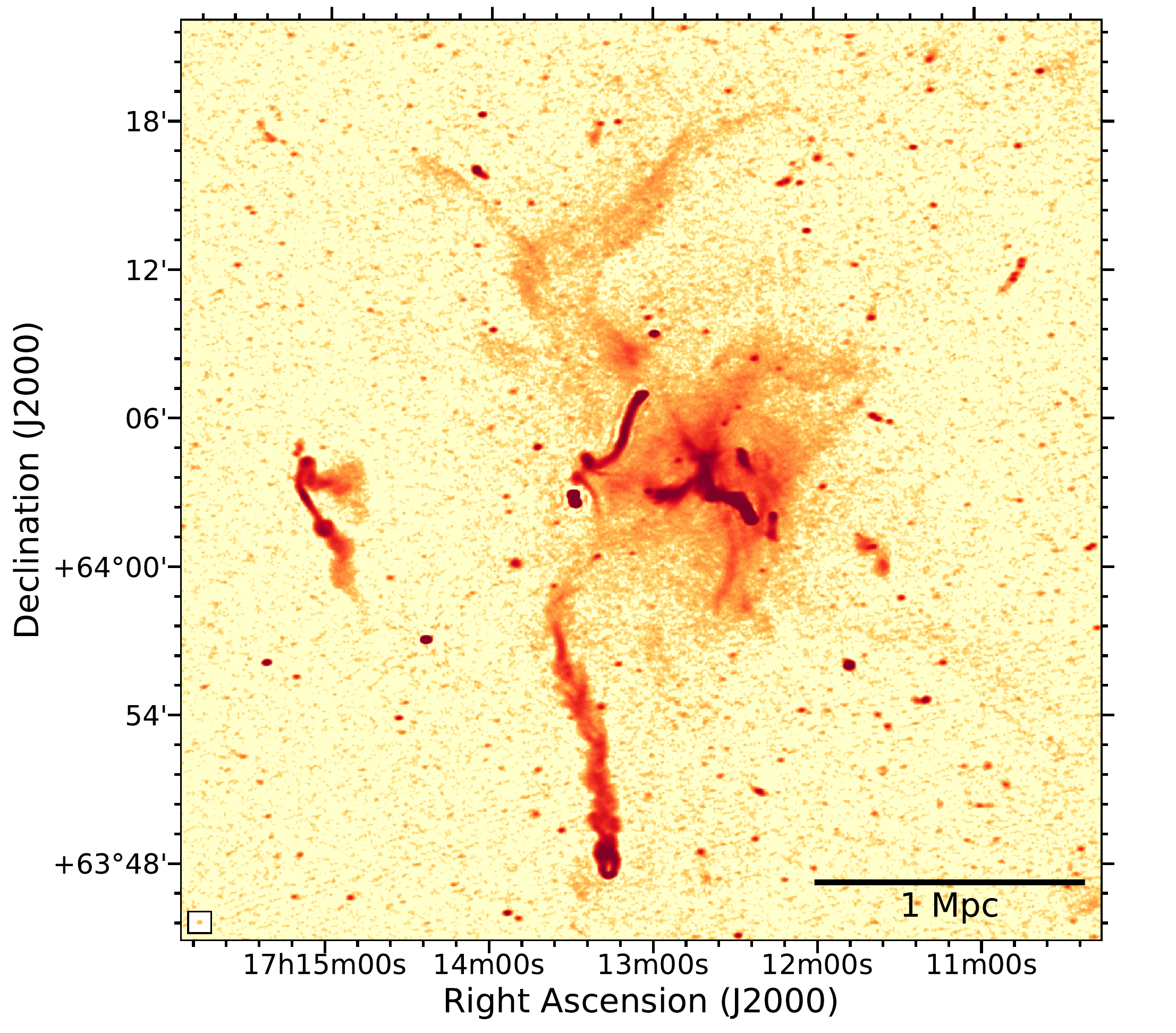}
 \caption{\lofar\ medium-resolution ($10.9 \arcsec \times 7.0 \arcsec$) image at 144 MHz of A2255 with $\sigma=85$ \mujyb obtained with \texttt{robust=-0.5} weighting of the visibilities and a Gaussian \uv-taper of 5\arcsec.}
 \label{fig:a2255_central}
\end{figure*}

\subsection{\wsrt}

We used the \wsrt\ 1.2~GHz dataset and image that were originally published by \citet{pizzo09} to study the spectral index properties of A2255. A full description of the deep observations ($4 \times 12$ hr with 4 different configurations of the \wsrt) and data reduction is given by \citet{pizzo09}. The final \wsrt\ image has a resolution of $15\arcsec\times14\arcsec$ and a noise level of 10 \mujyb.

\subsection{Chandra}

A2255 was observed twice with \chandra\ \acisi, for a total exposure time of 45 ks. For this work, we retrieved the observation with the longest exposure (40 ks, \obsid\ 894) available from the \chandra\ data archive. The dataset was processed with \ciao\ v4.11 from the \texttt{level=1} event file using the \chandra\ \caldb\ v4.8.2 following the standard \chandra\ data reduction threads\footnote{\url{http://cxc.harvard.edu/ciao/threads/index.html}}. Observation periods affected by soft proton flares were removed using the \texttt{lc\_clean} routine by inspecting the light curve extracted from the S2 chip in the $0.5-7.0$ \kev\ band. The final cluster image was created in the $0.5-2.0$ \kev\ band and has a cleaned exposure time of 32 ks. X-ray point sources were detected using the \texttt{wavdetect} task, confirmed by eye, and replaced with pixel values interpolated from surrounding regions with \texttt{dmfilth}. 

\begin{figure*}[t]
 \centering
 \includegraphics[width=\hsize,trim={0cm 0cm 0cm 0cm},clip]{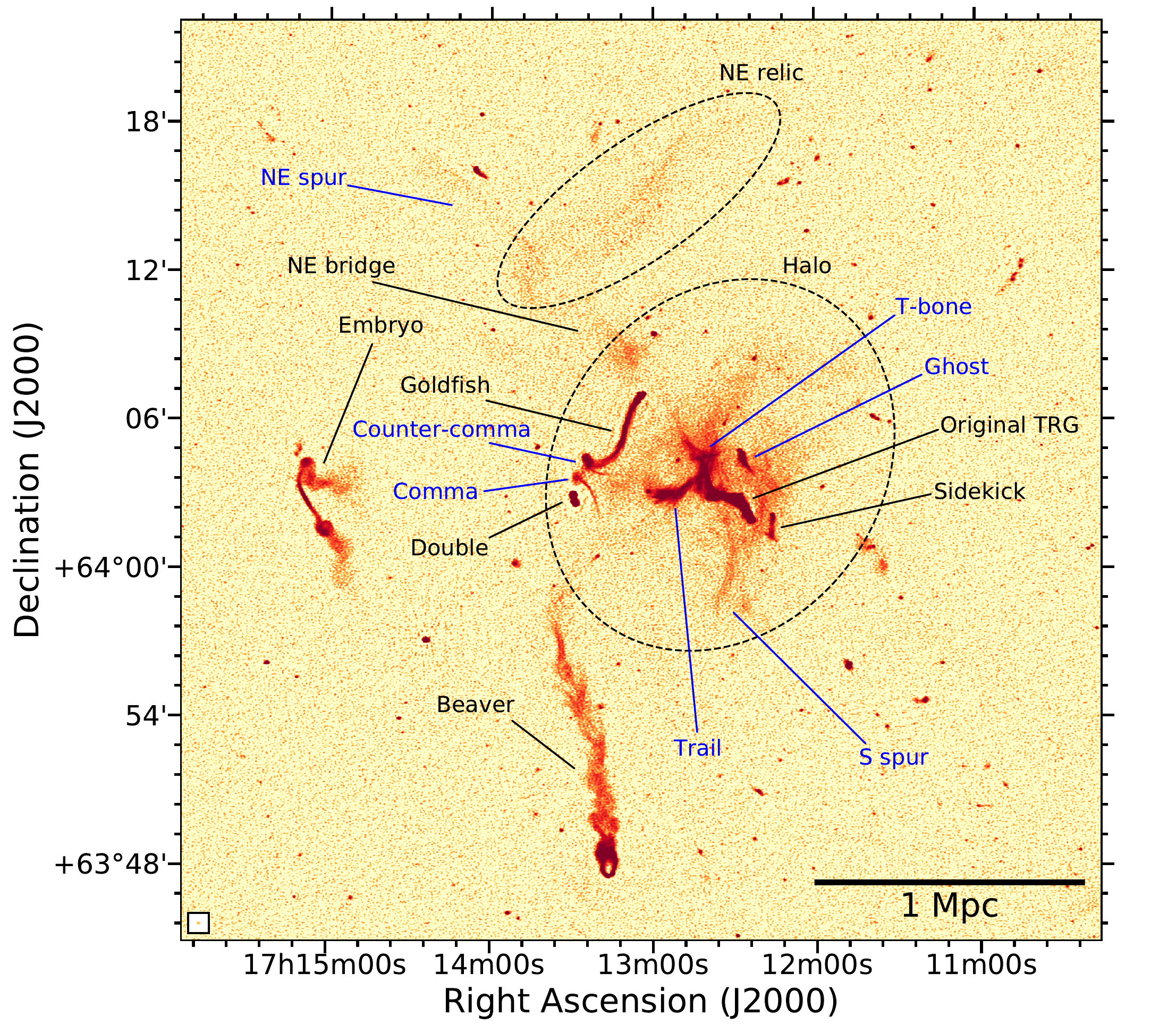}
 \caption{\lofar\ high-resolution ($5.0 \arcsec \times 3.8 \arcsec$) image at 144 MHz of A2255 with $\sigma=95$ \mujyb obtained with \texttt{robust=-1.0} (\ie\ close to uniform weighting) in which radio sources labeled in black follow the nomenclature used in \citet{harris80eight} while that labeled in blue are identified in this work for the first time.}
 \label{fig:a2255_highres}
\end{figure*}

Spectral analysis of the cluster thermal gas was performed with \xspec\ v12.10.0c \citep{arnaud96}. Contaminating radio and X-ray sources were removed from the \chandra\ event file before spectral extraction. We made use of blank-sky field datasets scaled by the ratio of the $9.5-12.0$ keV count rates to create background spectra. The ICM emission was modeled with an absorbed thermal plasma. The redshift of A2255 ($z=0.0806$), Galactic column density towards the cluster ($\nh = 2.74 \times 10^{20}$ cm$^{-2}$, from \citealt{willingale13}), and ICM metallicity (assumed 0.3 of the Solar value) were fixed during the fitting procedure. The normalization of the thermal model and the temperature were used to derive pseudo-pressure and pseudo-entropy values of the ICM following \citet{botteon18edges}.

\section{Results}\label{sec:results}

\subsection{The environment of A2255}

\begin{figure*}[t]
 \centering
 \includegraphics[width=.33\hsize,trim={0.9cm 0cm 0.9cm 0cm},clip]{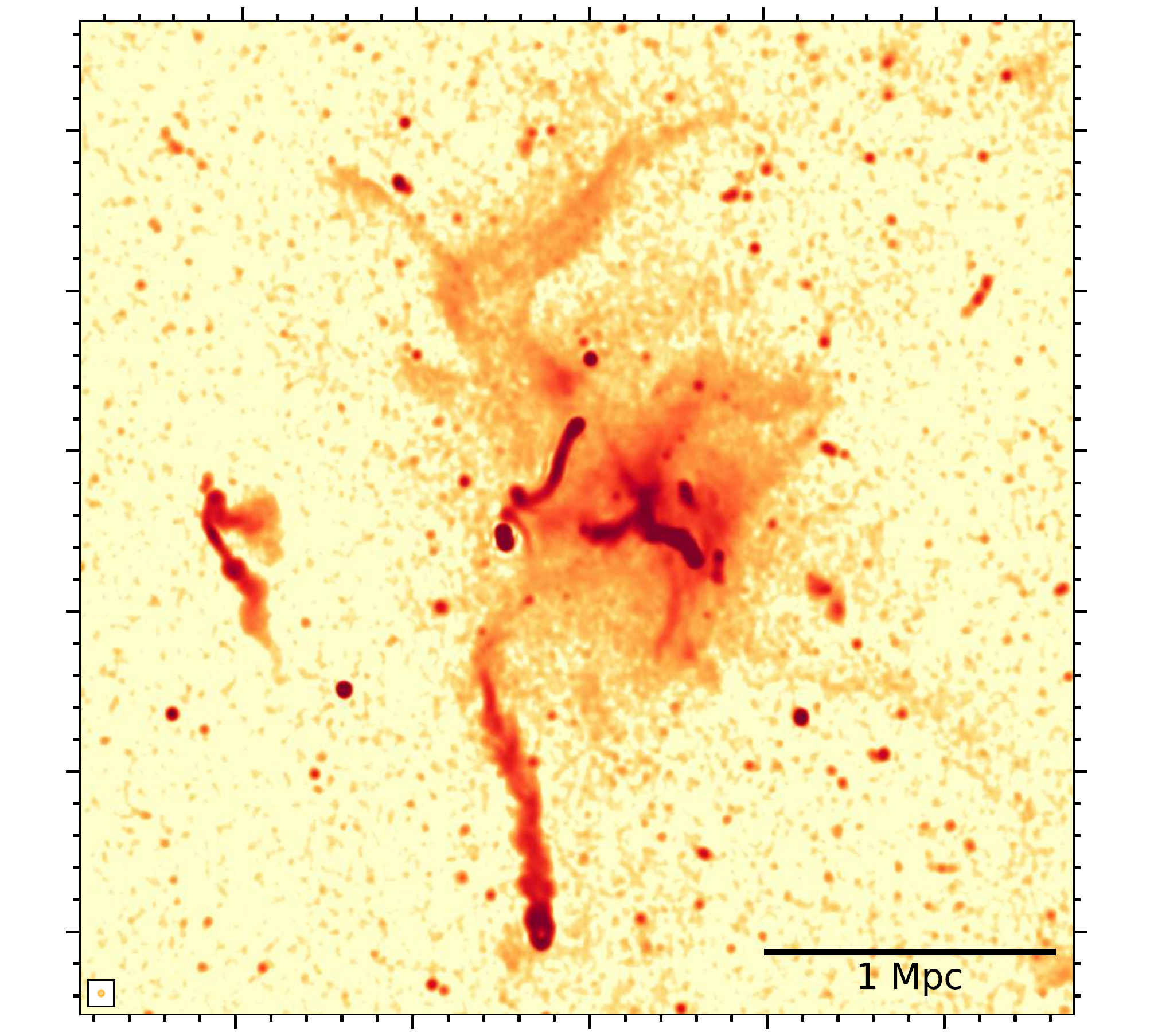}
 \includegraphics[width=.33\hsize,trim={0.9cm 0cm 0.9cm 0cm},clip]{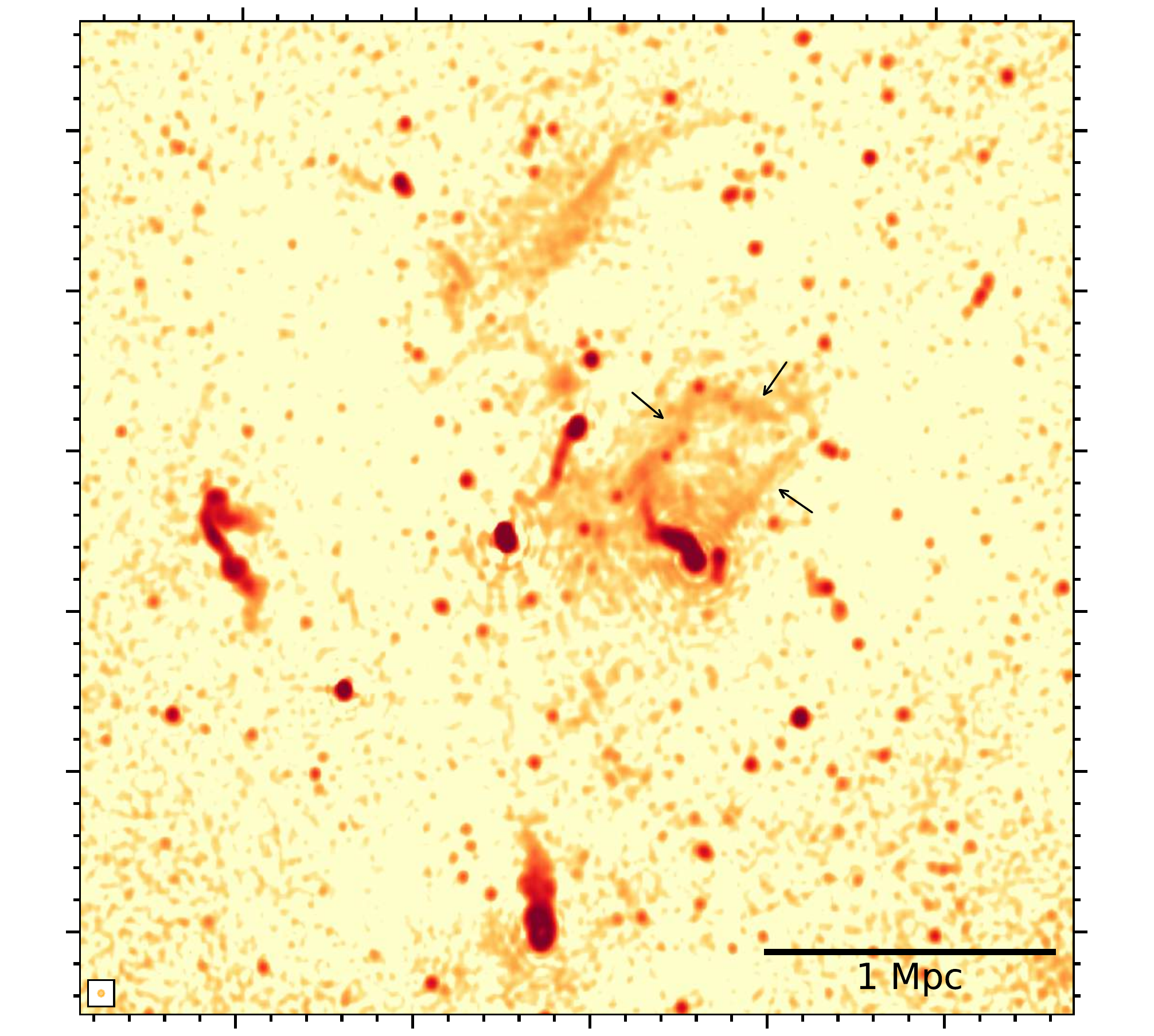}
 \includegraphics[width=.33\hsize,trim={0.9cm 0cm 0.9cm 0cm},clip]{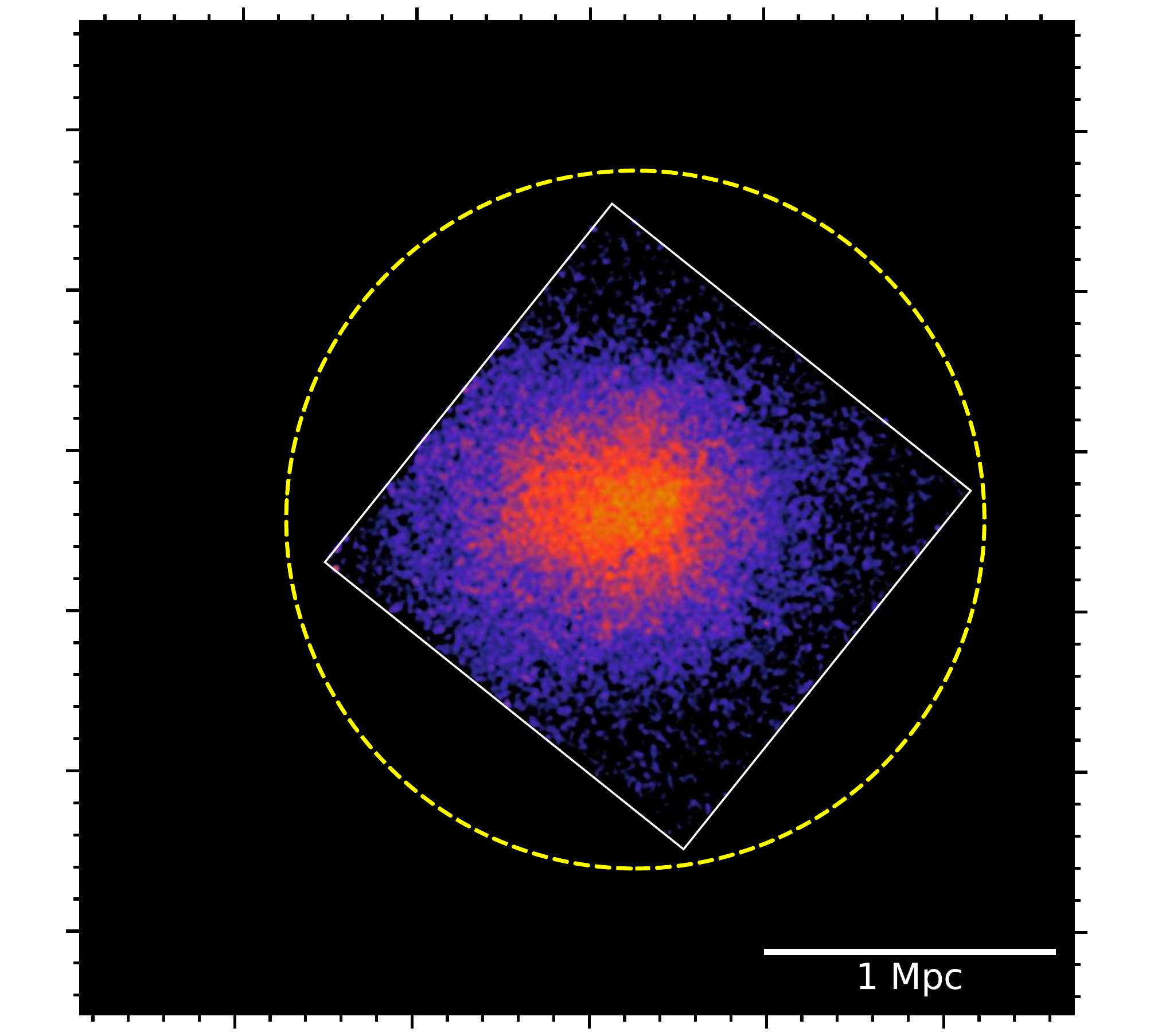}
 \caption{\lofar\ 144 MHz (\textit{left}) and \wsrt\ 1.2 GHz (\textit{center}, from \citealt{pizzo09}) images of A2255 at the same resolution of $15 \arcsec \times 14 \arcsec$. The noise levels in the images are $120$ \mujyb\ and 10 \mujyb\ for \lofar\ and \wsrt, respectively. The arrows in the \wsrt\ image indicate the three straight (polarized) filaments reported in \citet{govoni05} and \citet{pizzo11}. The \chandra\ image of the cluster in the $0.5-2.0$ keV band (\textit{right}) has been smoothed to a resolution of $\sim 5\arcsec$ for visualization purposes. The white region denotes the \acis-I FoV while the yellow circle indicates the approximate location of \rfive.}
 \label{fig:a2255_multifreq}
\end{figure*}

\lofar\ high sensitivity and high density of short baselines provide an unprecedented view on the multitude of radio structures in A2255 which range from a few tens of kpc up to cluster-scale emission. Remarkably, extended emission from the ICM is recovered by both the medium- ($10.9 \arcsec \times 7.0 \arcsec$) and high- ($5.0 \arcsec \times 3.8 \arcsec$) resolution images shown in Fig.~\ref{fig:a2255_central} and Fig.~\ref{fig:a2255_highres}. Reaching noise levels below 100 \mujyb, these are among the deepest images of a galaxy cluster ever produced so far at low-frequencies. This offers a unique opportunity to understand the origin of non-thermal phenomena in the ICM. \\
\indent
The FoV of the reported images spans $\sim10$ Mpc$^2$ centered on the radio halo. A number of bright sources are detected in this region, most of which are embedded in the diffuse emission. In Fig.~\ref{fig:a2255_highres}, we have labeled in black the sources following \citet{harris80eight}. These descriptive names were based on \wsrt\ observations at 610 MHz and 1.4 GHz. The highly sensitive and higher resolution \lofar\ images at 144 MHz show the presence of new sources in A2255 that we named and labeled in blue in Fig.~\ref{fig:a2255_highres} as well. Starting from the innermost region, the known sources are: the Goldfish, the Double, the Original Tailed Radio Galaxy (TRG), and the Sidekick. Each of these radio galaxies has an optical counterpart associated with the cluster \citep{miller03a2255}. The newly discovered sources are: the Trail, the T-bone, the Ghost, the Comma, and the Counter-comma. We shall show later that only the Trail seems directly related to a cluster member galaxy. These sources will be discussed in Section~\ref{sec:discussion_sb}. Diffuse emission in the ICM is found in the form of a radio halo (at the center) and a radio relic (in the NE). The two are connected by the NE bridge of radio emission. Two protuberances, the NE spur connected to the relic (tentatively reported by \citealt{pizzo09}) and the S spur embedded in the halo, are also detected. At $\sim 1.6$ Mpc from the cluster center, the extended radio galaxies known as the Embryo and the Beaver are observed. The tail of the Beaver extends over a Mpc, fading-out into the radio halo. \\
\indent
In this paper, we focus on the central ($\sim 1.5 \times 1.5$ Mpc$^2$), high signal-to-noise ratio (SNR) region of the cluster, with particular emphasis on the radio halo. The analysis of the emission on larger scales will be presented in a forthcoming paper that will exploit deeper (75~hr) \lofar\ observations.

\subsection{Surface brightness analysis}\label{sec:sb_ptp}

In Fig.~\ref{fig:a2255_multifreq} we report \lofar, \wsrt, and \chandra\ observations of A2255. The \lofar\ image (left) was convolved to the resolution of $15\arcsec\times14\arcsec$, which is the same resolution of the original \wsrt\ image (center) reported in \citet{pizzo09}. For visualization purposes, the \chandra\ image (right) was smoothed to a resolution of $\sim 5\arcsec$ and point sources were cosmetically removed as described above. The FoV of the \chandra\ observation covers the central region of the cluster, where the radio halo emission is observed with \wsrt\ at 1.2 GHz. \lofar\ recovers much more diffuse emission because (i) operating at low-frequency, it is more sensitive to steep spectrum emission and (ii) it has shorter baselines than the \wsrt\ 1.2 GHz observation. The radio halo extension is roughly twice in size at 144 MHz than at 1.2 GHz. In particular, compared to Fig.~\ref{fig:a2255_central}, the lower resolution enhances the faint emission in the region between the halo and NE relic, and between the end of the Beaver's tail and the Original TRG. \\
\indent
In the past, radio halos were typically described as smooth and regular sources with a morphology recalling that of the X-ray thermal emission. Highly sensitive observations with new generation instruments (\lofar, \ugmrt, \jvla) are changing our view of radio halos, unveiling the presence of a large diversity of surface brightness structures; A2255 is probably the most remarkable example. Its complex morphology with straight (polarized) filaments forming a rectangular shape (indicated by arrows in Fig.~\ref{fig:a2255_multifreq}, central panel) was already pointed out by $> 1$ GHz observations with the \vla\ \citep{govoni05} and \wsrt\ \citep{pizzo09, pizzo11}. The new \lofar\ data add further complexity to our picture of A2255, where a plethora of structures on various scales are embedded within the central diffuse emission.  

\begin{figure*}[ht]
 \centering
 \includegraphics[width=.33\hsize,trim={0cm 0cm 0cm 0cm},clip]{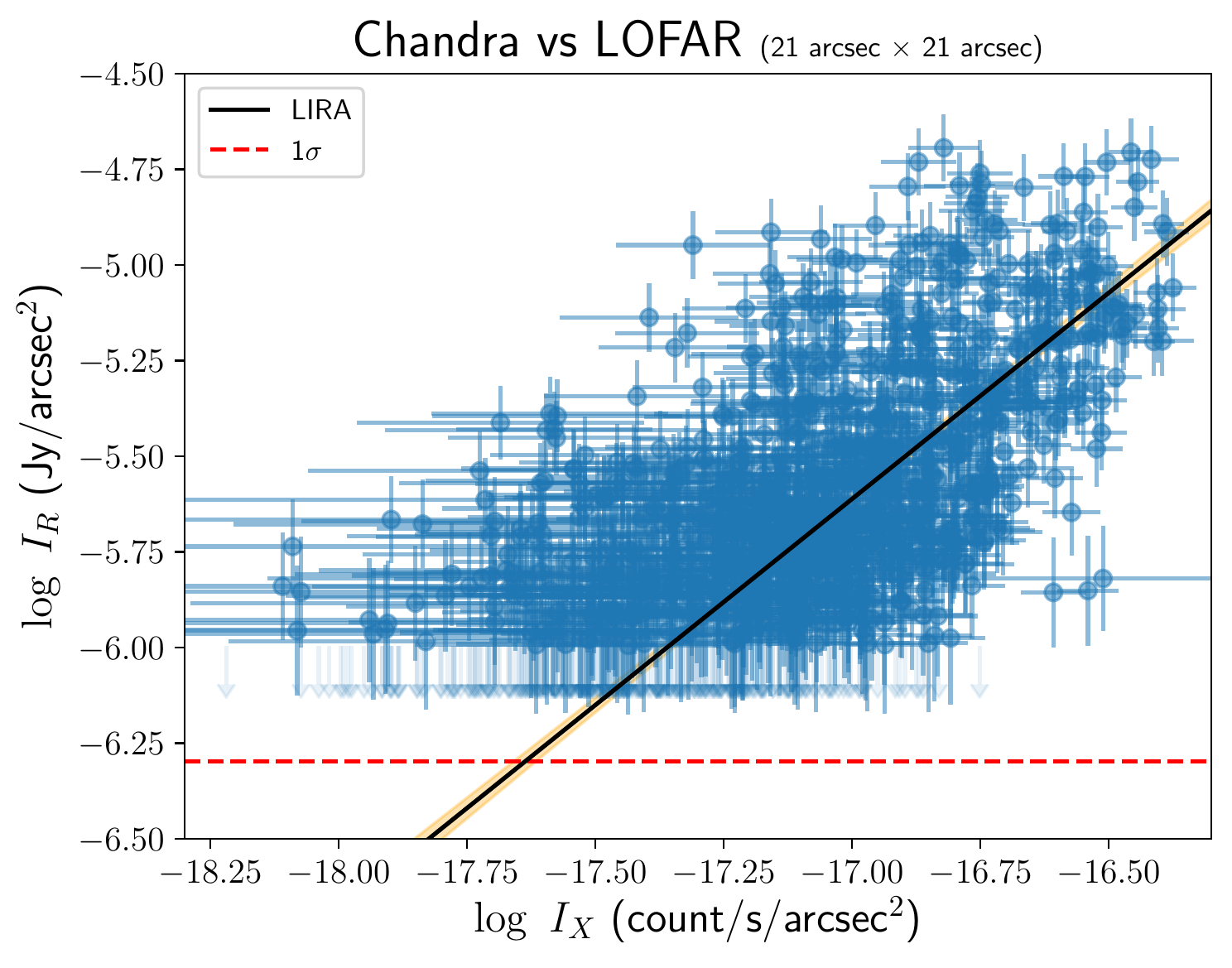}
 \includegraphics[width=.33\hsize,trim={0cm 0cm 0cm 0cm},clip]{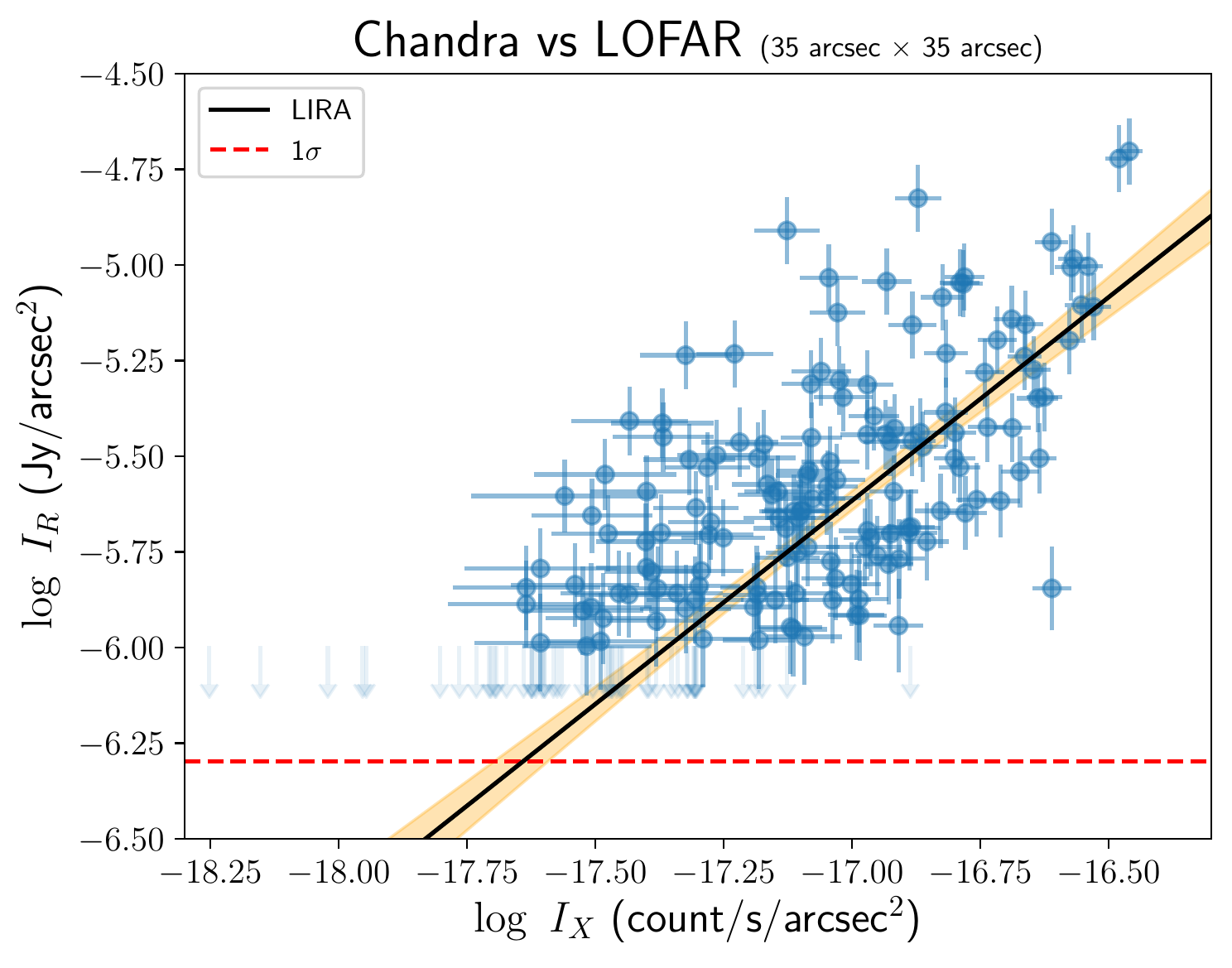}
 \includegraphics[width=.33\hsize,trim={0cm 0cm 0cm 0cm},clip]{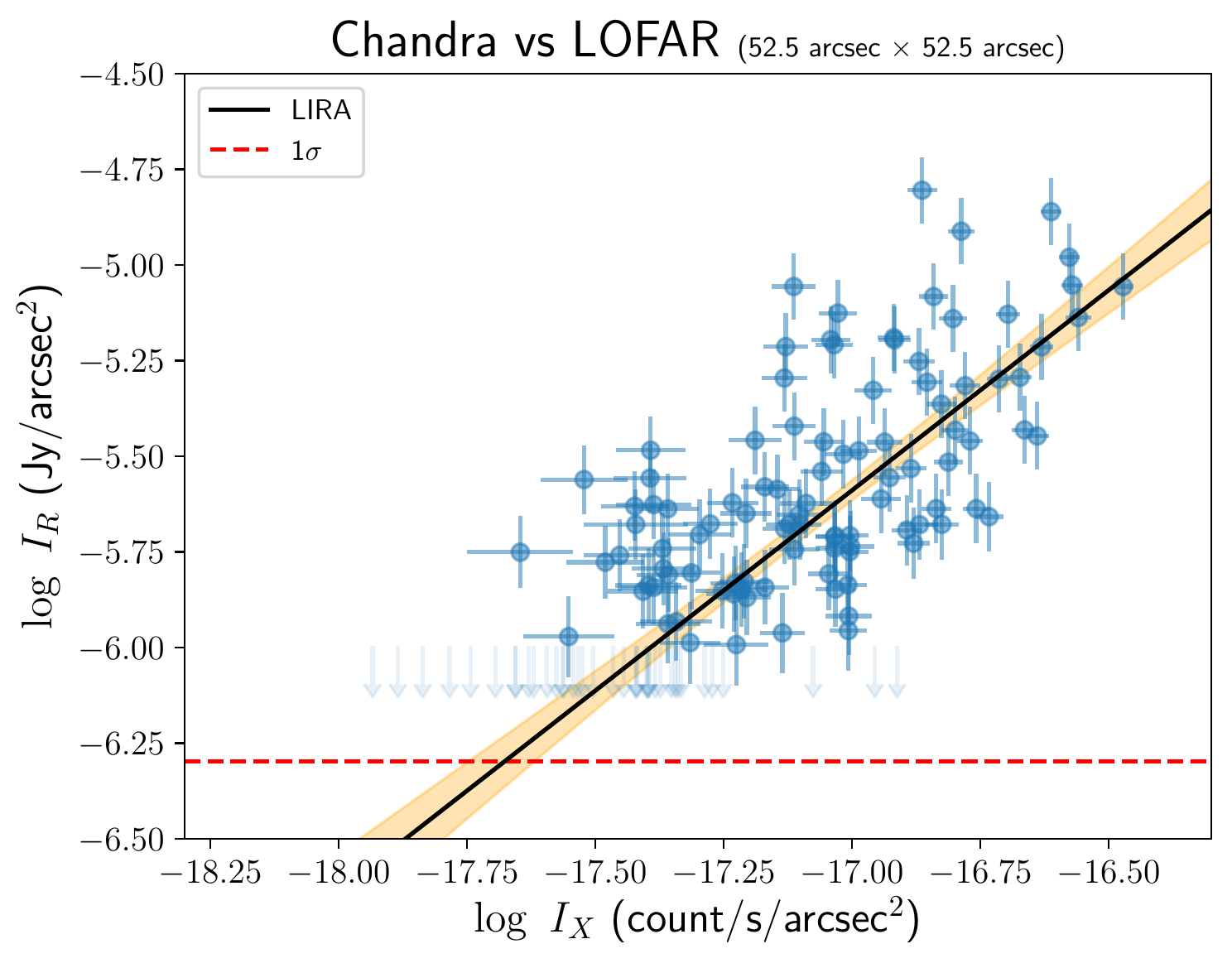}
 \caption{X-ray versus radio surface brightness computed on the same grid for the central region of A2255. The analysis was performed the \lofar\ image shown in Fig.~\ref{fig:a2255_multifreq} (left panel) using grids with different cell sizes (from \textit{left} to \textit{right}). Upper limits refer to cells where the radio surface brightness is below the $2\sigma$ level. Displayed errors on the radio measurements are obtained by summing in quadrature the statistic and systematic uncertainties. In the fitting process, only the statistical error was considered. \lira\ best-fit relations are reported with the corresponding 95\% confidence regions.}
 \label{fig:a2255_ptp_sb}
\end{figure*}

The close similarity between radio halos and clusters X-ray emission suggests an interplay between thermal and non-thermal components in the ICM  \citep[\eg][]{govoni01comparison}. This can be verified in a quantitative way by dividing the cluster into several square regions (or cells) that compose a grid where the X-ray and radio surface brightness are evaluated and compared. The data are generally described with a power-law of the generic form

\begin{equation}\label{eq:ptp_sb}
 \log(I_R) = A \log (I_X) + B
\end{equation}

\noindent
where the the slope of the scaling $A$ determines whether the radio brightness (\ie\ the magnetic field strength and relativistic particle density) declines faster (if $A>1$) than the X-ray brightness (\ie\ the thermal gas density), or vice versa (if $A<1$). \citet{govoni01comparison} found a slope consistent with a linear relation for A2255 by using \rosat\ and the \wsrt\ 333 MHz observations reported in \citet{feretti97}. \\
\indent
Thanks to the highly sensitive \lofar\ image at 144 MHz and the X-ray observation with higher angular resolution performed with \chandra, we can investigate the point-to-point radio/X-ray correlation in A2255 in a more detailed way. The analysis was performed on the images shown in Fig.~\ref{fig:a2255_multifreq} adopting the procedure described as follows. \\
\indent
We start with the creation of three grids. We selected a region of $14\times14$ arcmin$^2$ roughly centered on the radio halo (RA: 17h12m31.7s, DEC: $+$64\deg03\arcmin48.2\arcsec), and divided it into square cells with sizes $21\arcsec\times21\arcsec$, $35\arcsec\times35\arcsec$, and $52.5\arcsec\times52.5\arcsec$. The finest grid ensures that each cell is larger than the beam of the radio image. Cells outside the \chandra\ FoV are excluded from the grids. Since the goal is to investigate the connection between the non-thermal and thermal components in the ICM, we identified both point-like and extended (such as tailed radio galaxies) contaminating sources in the radio and X-ray images. We note that also the Trail and the T-bone (\cf\ Fig.~\ref{fig:a2255_highres}) were excluded because, despite their uncertain nature, their high brightness compared to the surrounding medium suggests a possible connection with radio galaxies. Cells intersecting the contaminating sources were removed from the grids. The final grids are reported in Appendix~\ref{app:grids}. Then, we calculated the integrated radio and X-ray surface brightness and compared them. The X-ray surface brightness values were obtained by converting the  \chandra\ \acisi\ count rate in the $0.5-2.0$ keV band in the cells assuming a thermal model with temperature as reported in Tab.~\ref{tab:properties}. Results are shown in Fig.~\ref{fig:a2255_ptp_sb}. We note the presence of a large intrinsic scatter in the plots, this may play a role in the determination of the slope between $I_R$ and $I_X$ (see below). 

\begin{table}[t]
 \centering
 \caption{Slope ($A$) from the \lira\ fitting to Eq.~\ref{eq:ptp_sb}, and Spearman ($r_s$) and Pearson ($r_p$) correlation coefficients of the data in Fig.~\ref{fig:a2255_ptp_sb}.}
 \label{tab:spearman_pearson}
  \begin{tabular}{lccc} 
  \hline
  \hline
  Parameters & $A$ & $r_s$ & $r_p$ \\
  \hline
  $I_R-I_X$ ($21\arcsec\times21\arcsec$) & $1.08\pm0.06$ & $0.58$ & $0.68$ \\
  $I_R-I_X$ ($35\arcsec\times35\arcsec$) & $1.06\pm0.15$ & $0.58$ & $0.61$ \\
  $I_R-I_X$ ($52.5\arcsec\times52.5\arcsec$) & $1.05\pm0.17$ & $0.63$ & $0.61$ \\
  \hline
  \end{tabular}
\end{table}

The two surface brightnesses are positively correlated, in agreement with other literature works \citep[\eg][]{govoni01comparison, feretti01, giacintucci05, rajpurohit18}. The general approach adopted in the past to find the slope of the correlation was to consider only data points above $3\sigma$ in the radio image during linear regression. The introduction of a threshold is indeed necessary as the reconstruction of an interferometric image can generate artifacts and biases leading to unreliable flux densities at low-SNR. However, we note that the introduction of a threshold combined with a large intrinsic scatter can introduce a bias in the correlation, if the selection effect is not taken into account by the regression method. In Appendix~\ref{app:threshold} we discuss these aspects in detail. In our case, we use a threshold of $2\sigma$ in the radio image and perform linear regression on Eq.~\ref{eq:ptp_sb} adopting the R-package \lira\footnote{\url{https://cran.r-project.org/web/packages/lira}} \citep{sereno16}, which is a Bayesian hierarchical approach to fit data with heteroscedastic and instrinsic scatter that can address selection effects in the response variable (Malmquist bias). In Tab.~\ref{tab:spearman_pearson} we summarize the slopes and correlation coefficients of the $I_R-I_X$ relations of Fig.~\ref{fig:a2255_ptp_sb}. Our results suggest a linear relation between radio and X-ray surface brightness.

\subsection{Spectral analysis}\label{sec:spectra_ptp}

\begin{figure*}[t]
 \centering
 \includegraphics[width=.49\hsize,trim={0cm 0cm 0cm 0cm},clip]{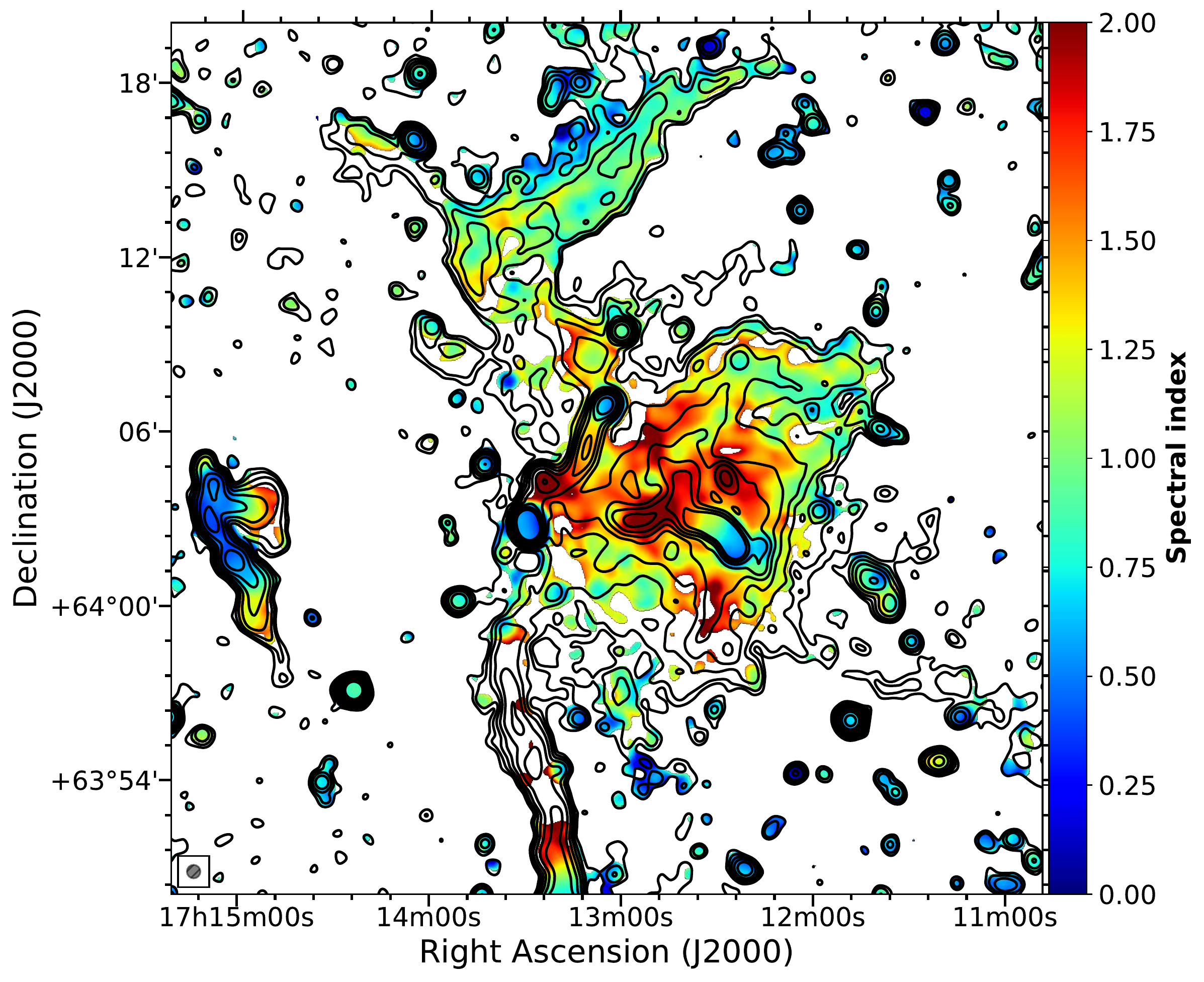}
 \includegraphics[width=.49\hsize,trim={0cm 0cm 0cm 0cm},clip]{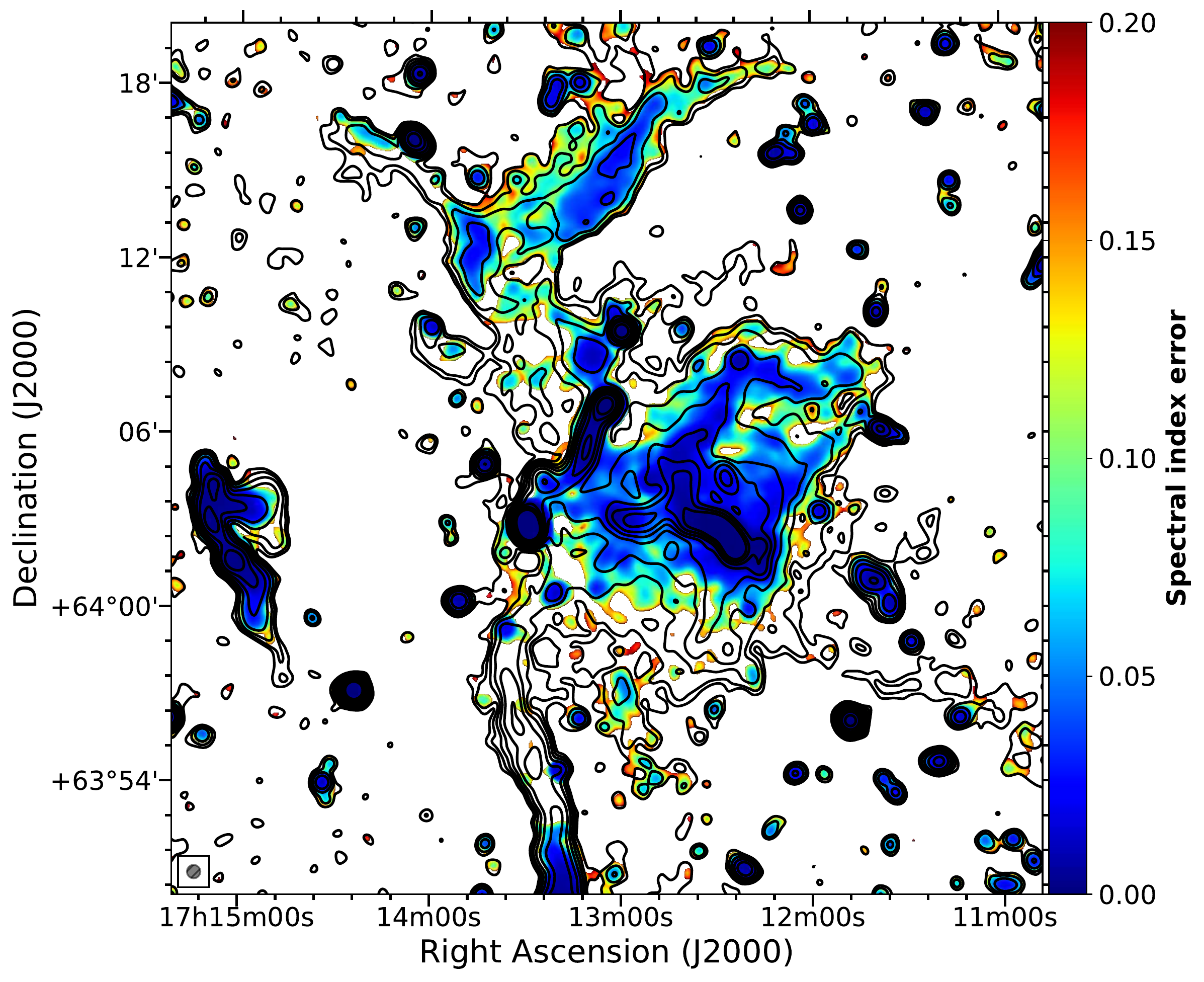}
 \caption{Spectral index map between 144 MHz and 1.2 GHz of the central radio emission in A2255 at a resolution of $28\arcsec\times28\arcsec$ (\textit{left}) with the corresponding (statistical) error map (\textit{right}). \lofar\ contours at the same resolution starting from $360$ \mujyb\ and spaced by a factor of 2 are shown. Pixels with values below $3\sigma$ were not considered in the computation of the spectral index.}
 \label{fig:spix}
\end{figure*}

Radio spectral index maps provide useful information on the origin of the diffuse and discrete synchrotron sources. We use the deep \lofar\ and \wsrt\ data to produce a spectral index map between 144 MHz and 1.2 GHz of the central region of A2255. The map, reported in Fig.~\ref{fig:spix}, has been obtained from images convolved to the same resolution of $28\arcsec\times28\arcsec$, regridded to identical pixel size, corrected for any position misalignment, and with matched inner-\uv\ coverage of $150\lambda$. The contribution of discrete sources (extended or point-like) can be easily disentangled from the diffuse emission of the ICM. Pixels with flux densities below $3\sigma$ in the radio images were blanked. Our spectral index map has more than a factor 5 higher resolution and covers a factor of $\sim2$ broader frequency range than those presented in \citet{pizzo09}. Although the adoption of a $3\sigma$ threshold is an approach commonly adopted in the literature for the computation of spectral index maps, we note that in our case this cut introduces a bias towards flatter spectral index values because the higher sensitivity of \lofar\ data to steep spectrum emission. The impact of different thresholds in the measurement of the spectral index and the its statistical distribution across the radio halo is investigated later in the paper.

By looking at Fig.~\ref{fig:spix}, nuclear emission from radio galaxies displays typical values $\alpha = 0.5-0.6$, while a signature of particle aging (\ie\ spectral steepening) is evident along the tails of the Original TRG, the Goldfish, and the Embryo. The most extreme case is represented by the Beaver, which is $\sim3$ times more extended at 144 MHz than at 1.2 GHz (Fig.~\ref{fig:a2255_multifreq}). By using low-resolution multi-frequency \wsrt\ observations, \citet{pizzo09} reported spectral index values up to $\alpha\sim3$ for the tail of the Beaver. \\
\indent
Radio relics often show clear radial spectral index gradients that are interpreted as electron aging due to merger shocks moving outwards \citep[\eg][]{giacintucci08, bonafede09double, vanweeren10}. The NE radio relic in A2255 manifests a marginal spectral steepening towards the cluster center (from $\alpha=0.8-0.9$ to $\alpha\sim1$) and it is characterized by two parallel structures elongated along the relic major axis that encompass a region with flatter spectrum ($\alpha=0.7-0.8$). These features are also observed in surface brightness (Fig.~\ref{fig:a2255_multifreq}) and are similar to the strands detected in the Toothbrush and Sausage relics \citep{rajpurohit18, digennaro18sausage}. Faint diffuse emission with flat spectrum ($\alpha=0.5-0.6$) is observed extending from the center of the relic towards the cluster outskirts in the putative pre-shock region (approximately at RA: 17h13m18s, DEC: $+$64\deg15\arcmin10\arcsec). This emission resembles that labeled as R5 in the Sausage relic \citep[see][]{digennaro18sausage}. The E region of the relic and the NE bridge below have comparable spectral index fluctuations with values up to $\alpha\sim1.4$. \\
\indent
The innermost region of A2255 shows a mix of radio plasma with a broad distribution of spectral index values. The steepest spectrum emission is provided by the newly discovered Ghost, T-bone, and Trail sources, which are characterized by $\alpha=2-2.5$. Despite the high number of discrete sources embedded in the radio halo, a complex spectral index distribution can also be inferred for the diffuse ICM emission. The flatter values ($\alpha\sim1$) are found in the NW region of the halo, which is clearly detected by \wsrt\ at 1.2 GHz (Fig.~\ref{fig:a2255_multifreq}, central panel). Patches of emission with very steep spectrum ($\alpha>1.6$) are located around the Ghost, between the Trail and the Goldfish, and in the S spur.

\begin{figure*}[t]
 \centering
 \includegraphics[width=.33\hsize,trim={0cm 0cm 0cm 0cm},clip]{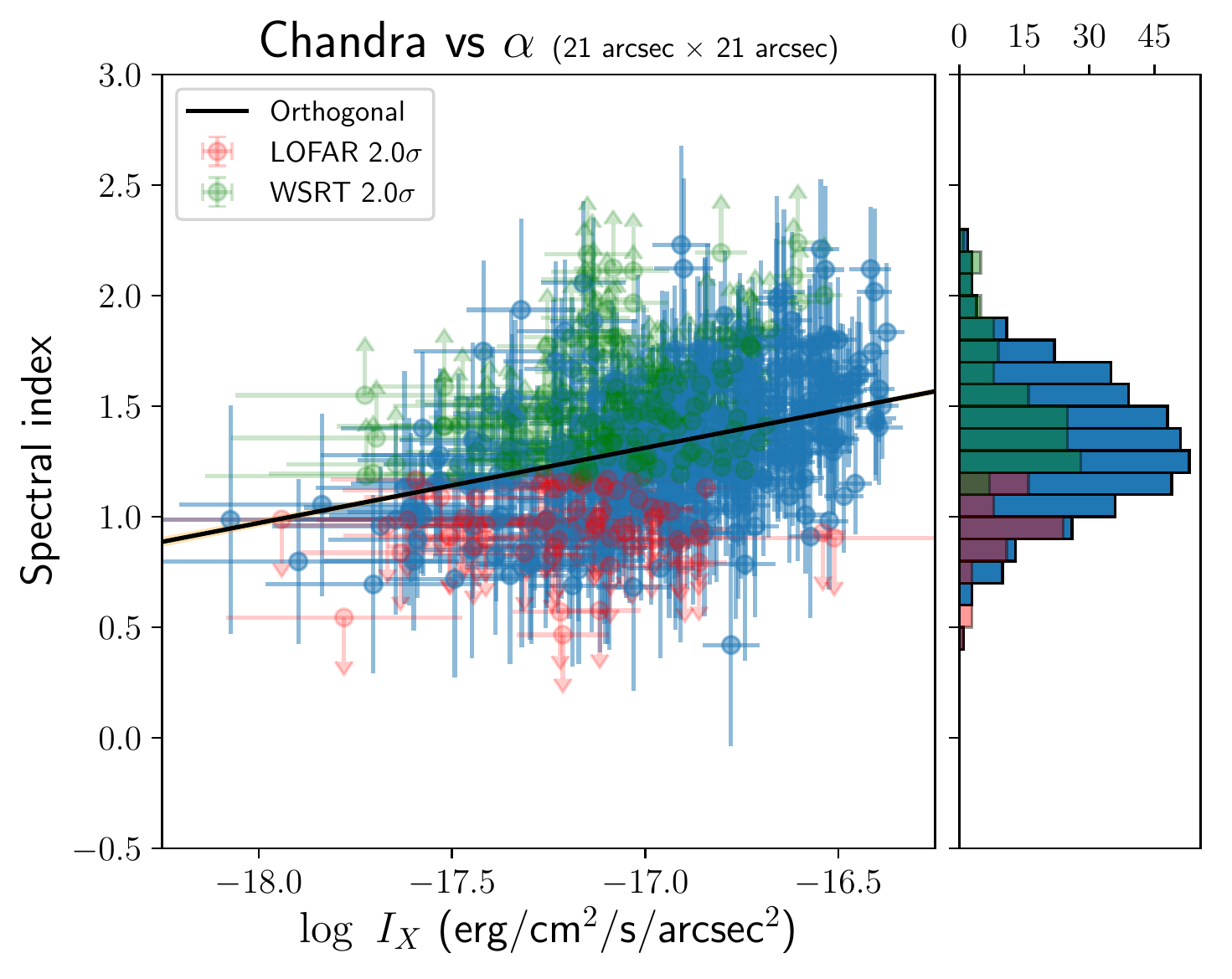}
 \includegraphics[width=.33\hsize,trim={0cm 0cm 0cm 0cm},clip]{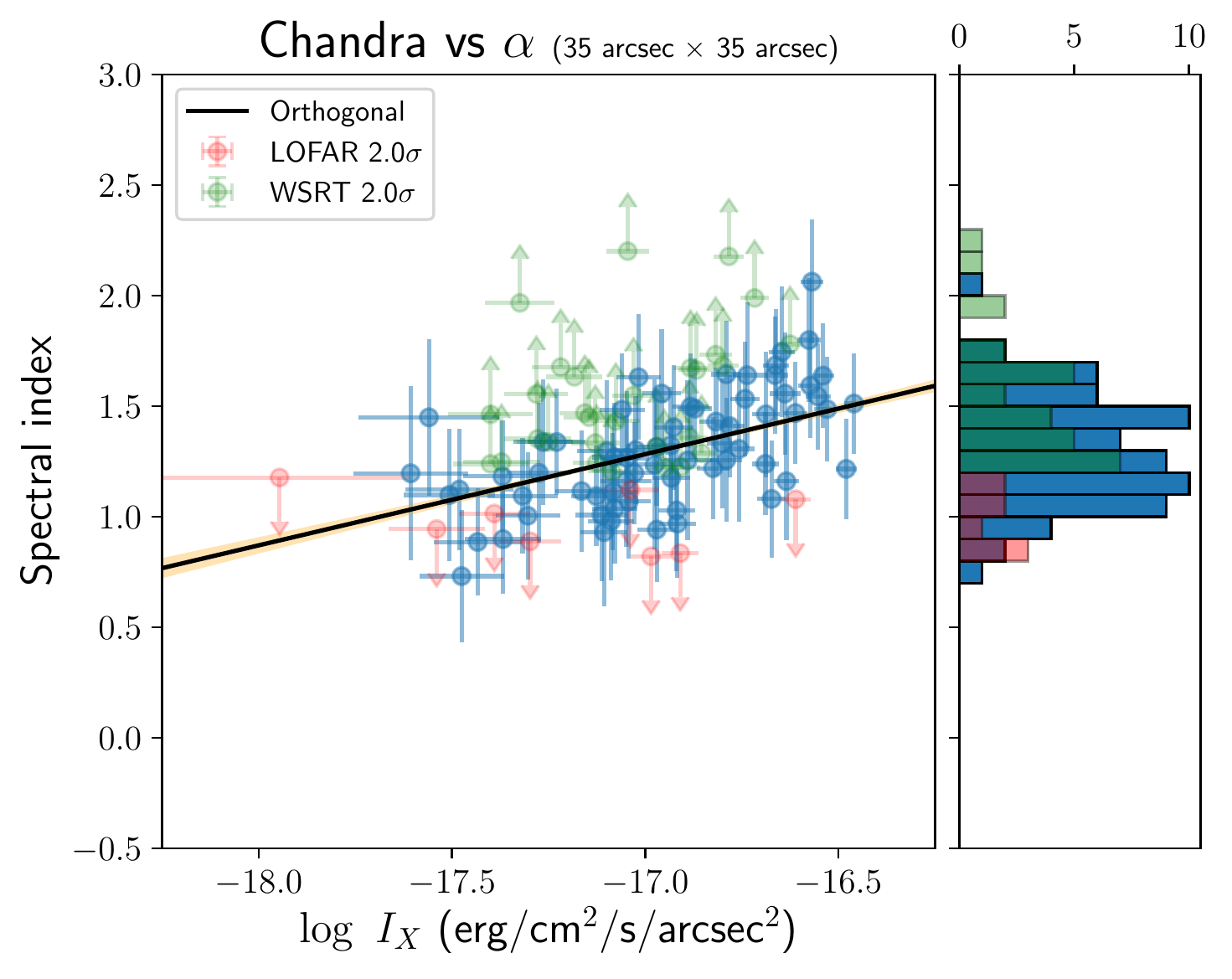}
 \includegraphics[width=.33\hsize,trim={0cm 0cm 0cm 0cm},clip]{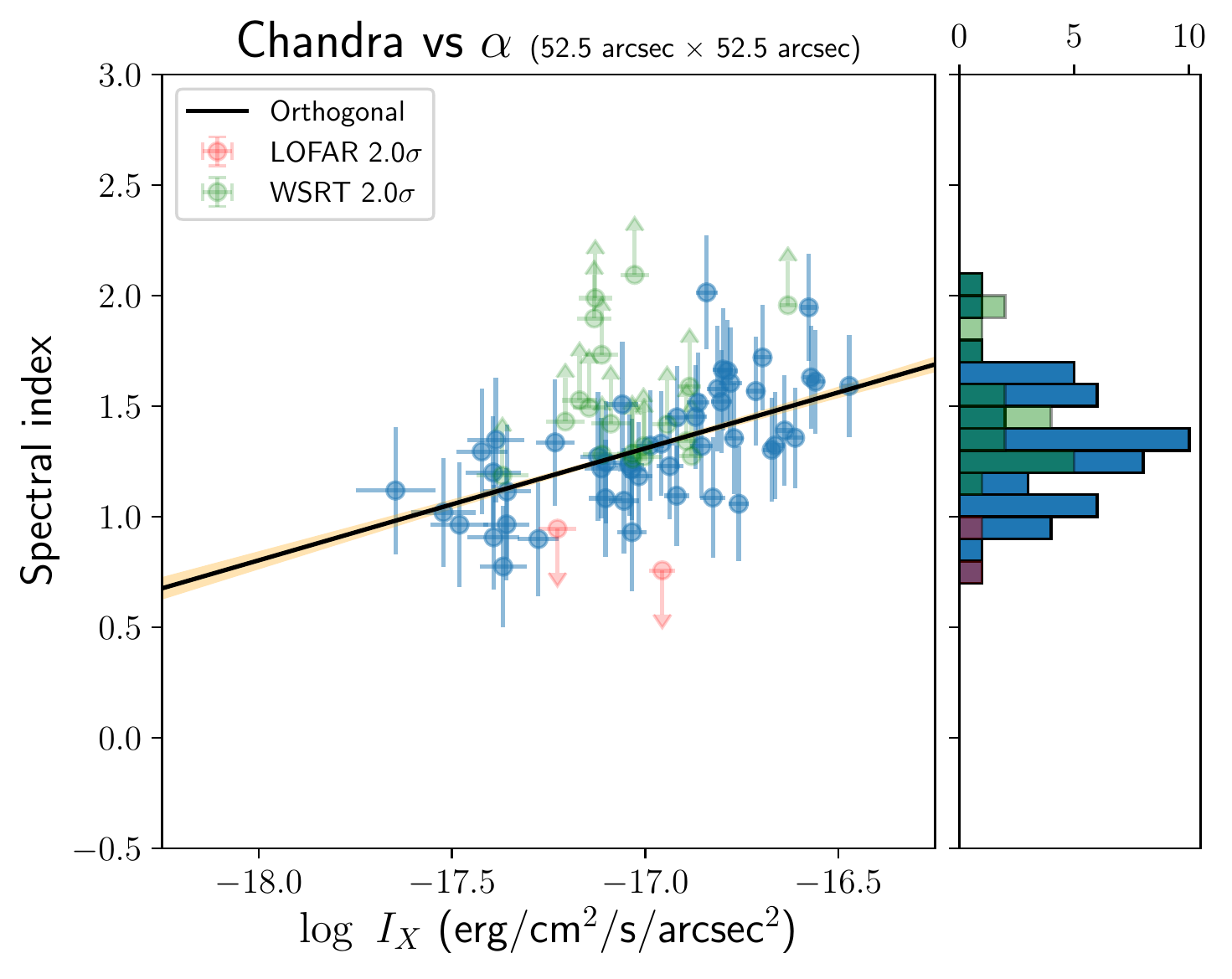}
 \caption{X-ray surface brightness versus spectral index of the radio halo emission (in blue) using grids with different cell sizes (from \textit{left} to \textit{right}). The histograms of spectral index values are shown on the right panels. The \lofar\ and \wsrt\ $2\sigma$ limits are reported in red and green, respectively. Orthogonal-BCES best-fit relations are reported with the corresponding 95\% confidence regions.}
 \label{fig:alpha_histo}
\end{figure*}

We study more quantitatively the distribution of spectral index values in the halo and its connection with the thermal gas with a point-to-point analysis on the images shown in Fig.~\ref{fig:a2255_multifreq} (note that we produced another \lofar\ image at $15\arcsec\times14\arcsec$ with $\uv_{min}=150\lambda$ for this kind of analysis). We follow a similar approach to that described in Section~\ref{sec:sb_ptp}, rejecting the same contaminating discrete sources and adopting the three grids characterized by different cell sizes. As discussed in Appendix~\ref{app:threshold}, only cells with emission above $2\sigma$ in the radio images are considered in the computation of $\alpha$ and in the calculation of the correspondent X-ray surface brightness. Instead, cells where the emission is below this threshold in one of the two instruments are considered as upper or lower limits on the spectral index assuming $I_R = 2\sigma$. Results are shown in Fig.~\ref{fig:alpha_histo}. In Appendix~\ref{app:threshold} we investigate how different thresholds impact this kind of analysis. \\
\indent
The right panels of Fig.~\ref{fig:alpha_histo} show the histograms of the spectral index values across the radio halo. We found that the distribution mean value and its standard deviation have small variation with the adopted grid and thresholds, being $\overline{\alpha} \sim 1.3$ and $\sigma_\alpha \sim 0.3$ (\cf\ Tab.~\ref{tab:threshold} in Appendix~\ref{app:threshold}). Following the argument of \citet{vacca14} and \citet{pearce17}, if the broad fluctuation of $\alpha$ is due to due to measurement errors, the mean value of the spectral index error should be comparable to the standard deviation of the spectral index distribution. We found that $\sigma_\alpha$ is 3 times larger than the (statistical) measurement error, indicating that the dispersion of spectral index values is statistically significant in the radio halo in A2255. We note that the mean spectral index obtained is in good agreement with the integrated spectral index of $\alpha \sim 1.3$ reported in \citet{pizzo09}. Nonetheless, the larger number of lower limits compared to number of upper limits in Fig.~\ref{fig:alpha_histo} indicates the presence of a bias towards flatter spectral indexes due to the limited sensitivity of the \wsrt\ data. Therefore, the average spectrum is likely steeper than $\overline{\alpha}=1.3$ and the dispersion is probably also larger than $\sigma_\alpha=0.3$. Deeper observations at high frequency are required to confirm this statement. \\
\indent
Fig.~\ref{fig:alpha_histo} also shows a hint of a correlation between the spectral index of the radio halo and the X-ray surface brightness of the ICM. The Spearman and Pearson correlation coefficients and the best-fit slopes obtained assuming a relation in the form

\begin{equation}\label{eq:ptp_alpha}
 \alpha = A \log (I_X) + B
\end{equation}

\noindent
are summarized in Tab.~\ref{tab:threshold} in Appendix~\ref{app:threshold}. A few words of caution on the slopes obtained should be given. Contrary to the case of the $I_R - I_X$ relation, where the distribution of data points has a clear cut on the $y$-axis (Fig.~\ref{fig:a2255_ptp_sb}), the $\alpha-I_X$ plane is populated both by upper and lower limits at different levels. These generate a complex regression problem that can not be handled even by a sophisticated Bayesian method such as \lira. For this reason, we performed linear regression only taking into account the ``detections'' (blue points in Fig.~\ref{fig:alpha_histo}) using the BCES-orthogonal method. Although ignoring the presence of upper/lower limits could affect the determination of the slope, we note that it would not change the existence of a positive and mild trend between spectral index and X-ray surface brightness, which is the main result that we want to highlight. In this respect, the slopes reported in the paper serve only as an indication of the steepness of the correlation and should not be not used to draw physical conclusion. In Appendix~\ref{app:threshold} we report results for different threshold values.

\begin{figure}[t]
 \centering
 \includegraphics[width=\hsize,trim={0cm 0cm 0cm 0cm},clip]{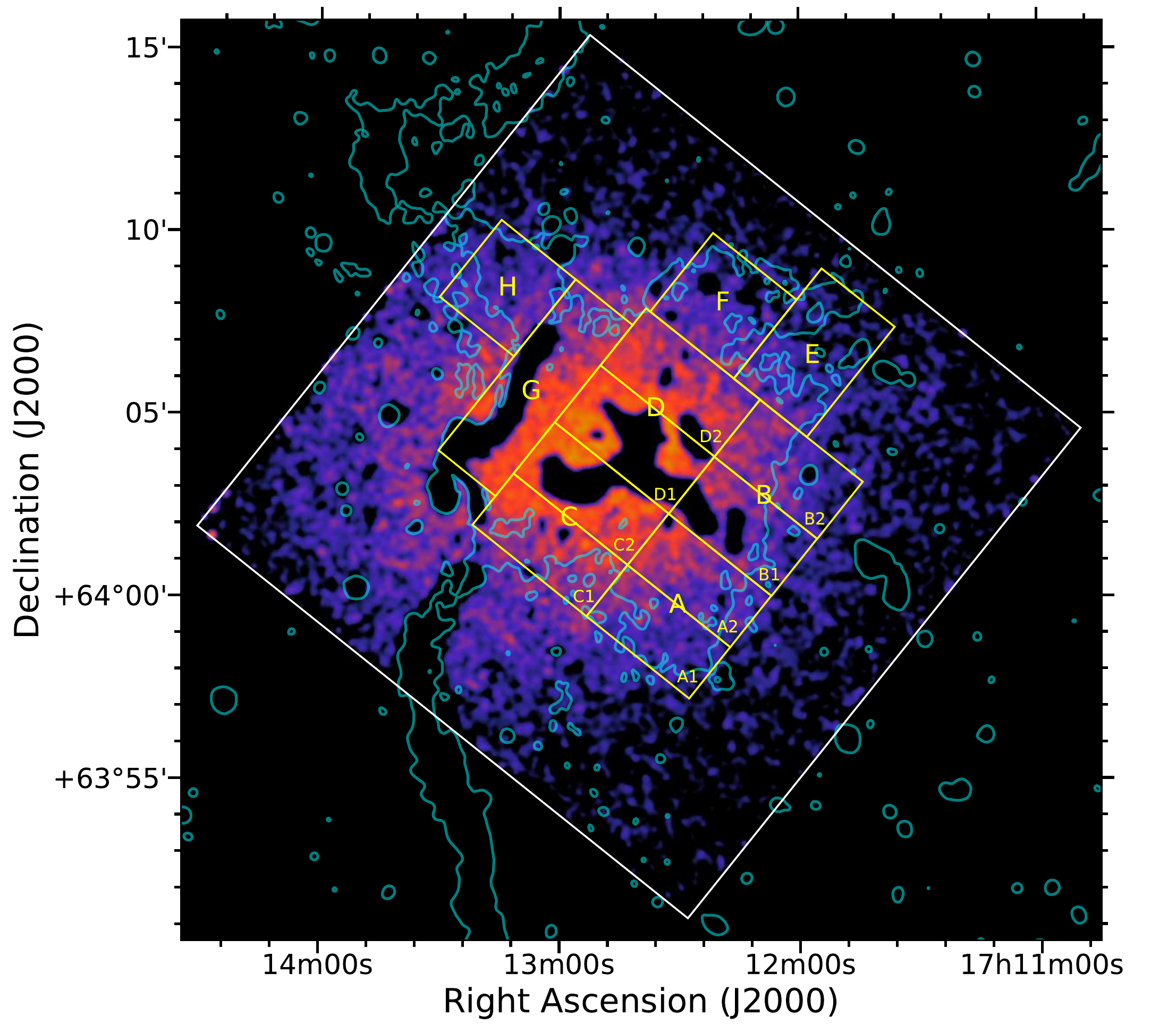}
 \caption{Regions adopted for the (radio and X-ray) spectral analysis and used to derive the quantities shown in Fig.~\ref{fig:alpha_thermo}. The $5\sigma$ level \lofar\ contour (Fig.~\ref{fig:a2255_multifreq}) is overlaid in cyan on the \chandra\ color image. Contaminating \lofar, \wsrt, and \chandra\ sources are blanked.}
 \label{fig:binmap}
\end{figure}

\begin{figure*}[t]
 \centering
 \includegraphics[width=.33\hsize,trim={0cm 0cm 0cm 0cm},clip]{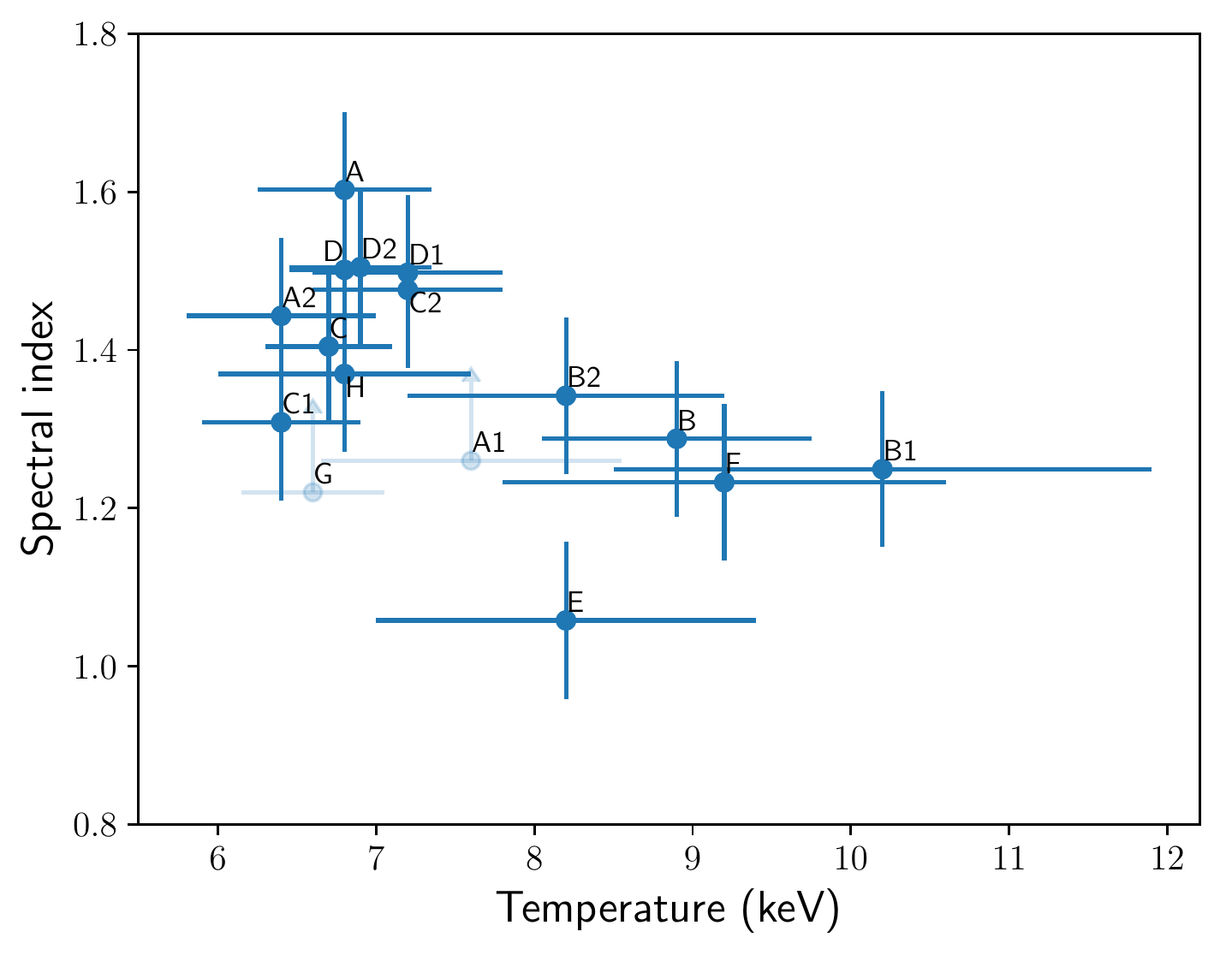}
 \includegraphics[width=.33\hsize,trim={0cm 0cm 0cm 0cm},clip]{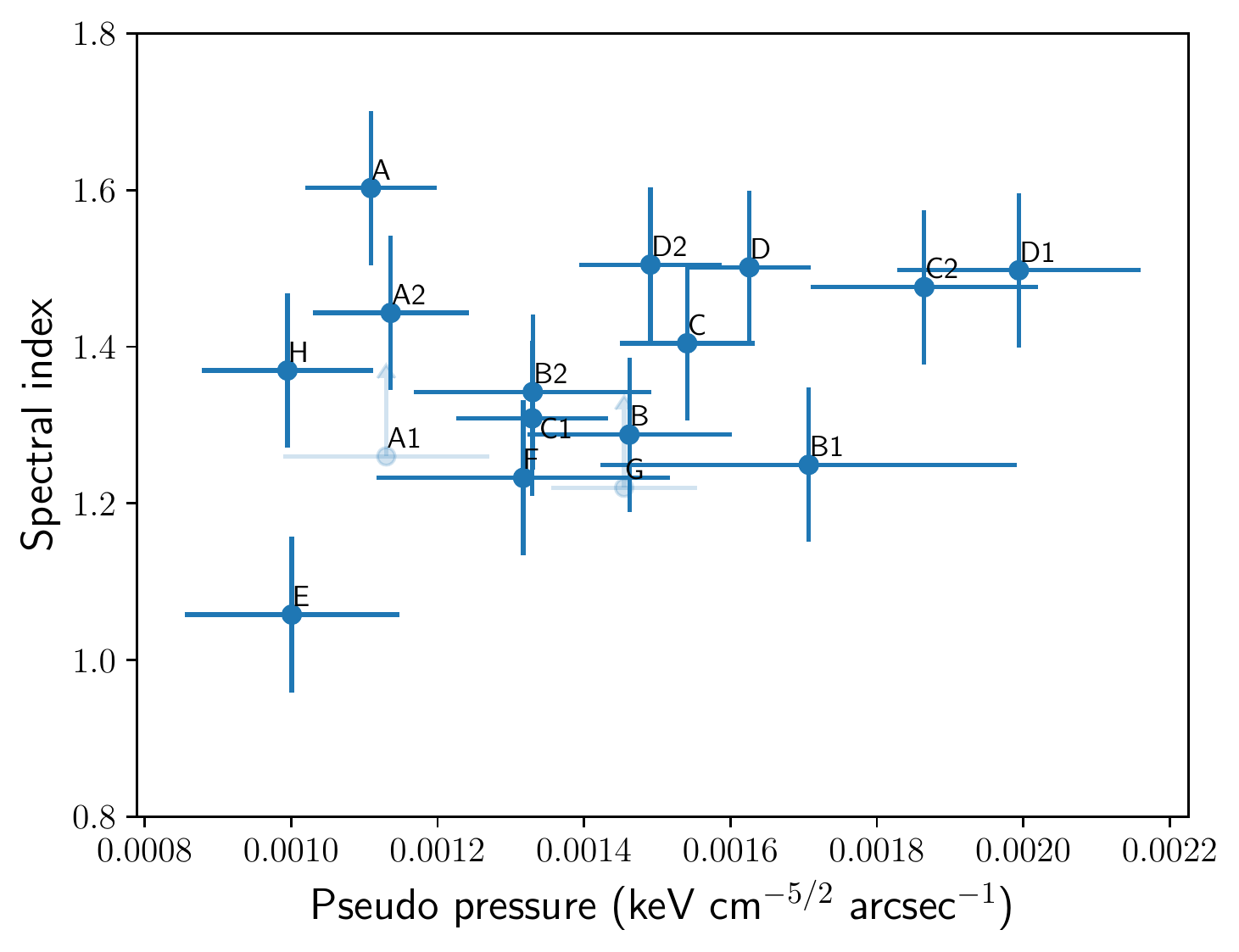}
 \includegraphics[width=.33\hsize,trim={0cm 0cm 0cm 0cm},clip]{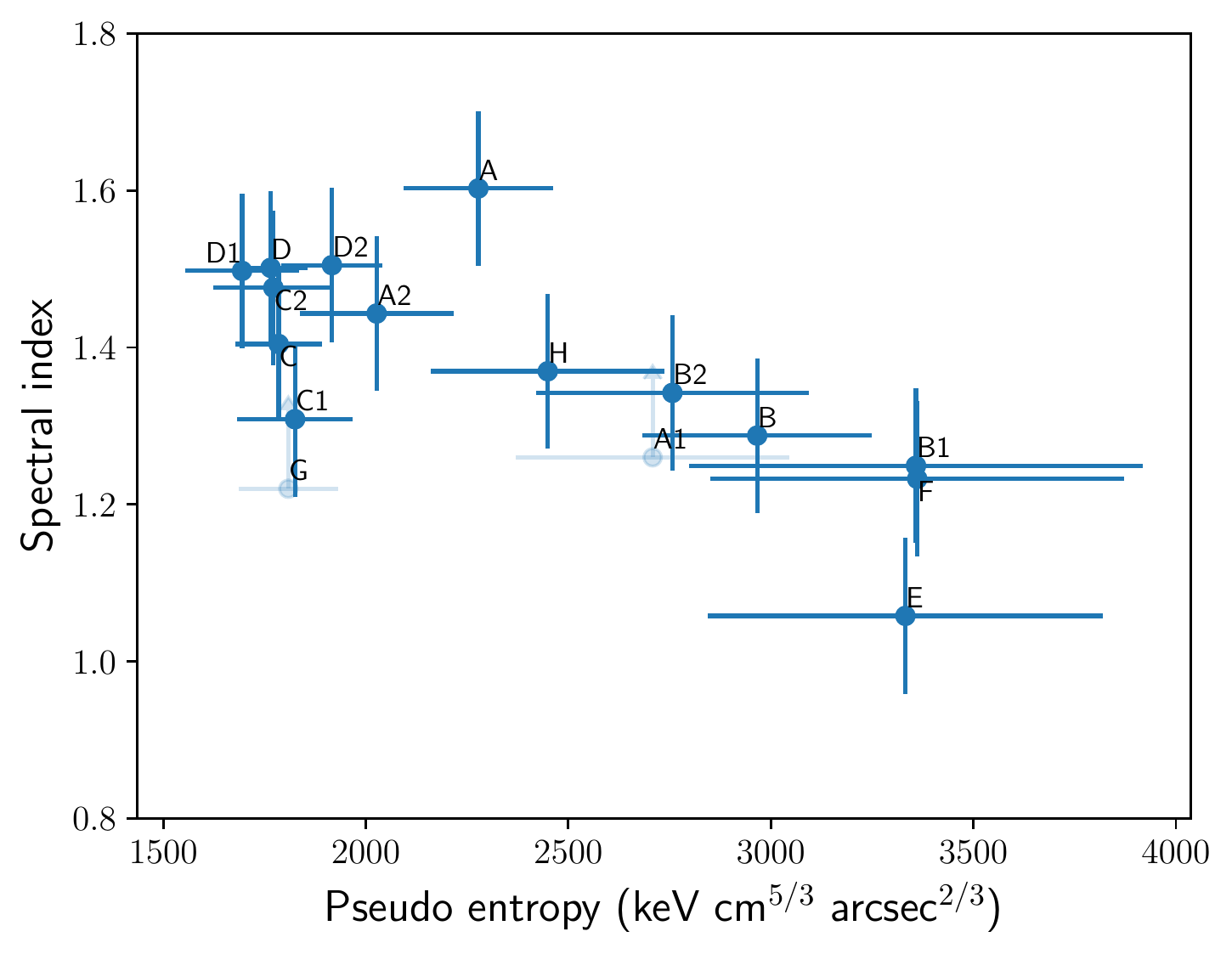}
  \caption{Spectral index of the radio halo against the thermodynamical quantities of the ICM evaluated in the same regions: temperature (\textit{left}), pseudo-pressure (\textit{center}), and pseudo-entropy (\textit{right}). The prefix pseudo- underlines that these quantities are projected along the line of sight. Lower limits on the spectral index have been computed assuming the $2\sigma$ level emission in \wsrt. Points are labeled following Fig.~\ref{fig:binmap}.}
 \label{fig:alpha_thermo}
\end{figure*}

To complete the spectral study of the cluster, we extract and fit spectra of the thermal ICM and compare the results with the values of the spectral index of the diffuse radio emission evaluated in the same regions. The selection of the regions is not trivial as we have to deal with three datasets (\lofar, \wsrt, and \chandra) characterized by different SNR. We roughly followed the \lofar\ $5\sigma$ level emission and divided the cluster into 8 regions labeled from A to H in Fig.~\ref{fig:binmap}. The largest and innermost regions A-D were further divided into two additional subregions. In Fig.~\ref{fig:alpha_thermo} we compare the radio spectral index with the thermodynamical quantities of the ICM. Due to the small number of points, we do not perform linear regression and we just report the Spearman and Pearson correlation coefficients in Tab.~\ref{tab:spearman_pearson_thermo}. Whereas there is not strong evidence for tight relations between spectral index and temperature/pseudo-pressure (the Spearman and Pearson coefficients vary in the range $\sim|0.2-0.8|$), the presence of a negative relation between $\alpha$ and pseudo-entropy seems to hold even when the uncertainties on $r_s$ and $r_p$ are taken into account.

\begin{table}[t]
 \centering
 \caption{Spearman ($r_s$) and Pearson ($r_p$) correlation coefficients for the points corresponding to subregions A-D, E, F, and H in Fig.~\ref{fig:alpha_thermo}. Due to the low number of data points, we evaluate the uncertainties in the coefficient via $10^4$ bootstrap realizations of the correlations by shuffling the data points with repetition.}
 \label{tab:spearman_pearson_thermo}
  \begin{tabular}{lcc} 
  \hline
  \hline
  Parameters & $r_s$ & $r_p$ \\
  \hline
  $kT-\alpha$ & $-0.51\pm0.28$ & $-0.57\pm0.19$ \\
  $P-\alpha$ & $+0.49\pm0.30$ & $+0.51\pm0.27$ \\
  $K-\alpha$ & $-0.78\pm0.19$ & $-0.84\pm0.12$ \\
  \hline
  \end{tabular}
\end{table}

\section{Discussion}\label{sec:discussion}

The new \lofar\ images at 144 MHz of A2255 show a high level of complexity of the emission over a broad range of spatial scales. The spectral analysis in combination with the published \wsrt\ data at 1.2 GHz \citep{pizzo09} reveals a mixture of electrons with different properties in the cluster environment, with regions characterized by ultra-steep spectrum ($\alpha>2$) emission. A complicated system such as A2255 deserves multiple studies ranging from the population of radio galaxies with peculiar morphologies up to the extended diffuse emission in the cluster outskirts. Due to the forthcoming deep \lofar\ observations (75~hr) of the cluster, in this paper we focused on the radio halo and investigated its connection with the thermal gas. In what follows, we discuss our results and compare them with those from other radio halos.

\subsection{Surface brightness properties}\label{sec:discussion_sb}

\begin{figure}[ht]
 \centering
 \includegraphics[width=\hsize,trim={0cm 0cm 0cm 0cm},clip]{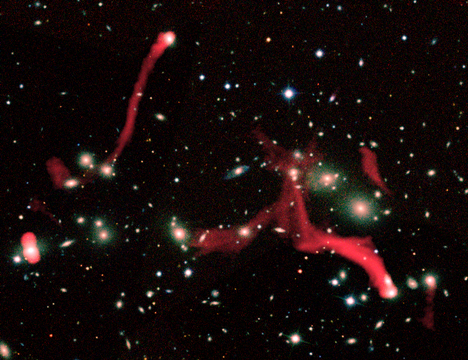}
 \caption{Composite optical (\sdss, RGB) + radio (\lofar\ 144 MHz, red) image of the central region of A2255. The Double, the Goldfish, the Trail, the T-bone, the Original TRG, the Ghost, and the Sidekick radio sources are displayed (\cf\ Fig.~\ref{fig:a2255_highres}). The FoV of the panel is $\sim$ 0.9 Mpc $\times$ 0.7 Mpc.}
 \label{fig:opt-radio}
\end{figure}

One of the most remarkable features that emerges from the \lofar\ images (Fig.~\ref{fig:a2255_central} and Fig.~\ref{fig:a2255_highres}) is the multitude of interactions between radio galaxies, the halo, and the ICM in the cluster center. In Fig.~\ref{fig:opt-radio} we show an optical/radio overlay of the central region of A2255. Clear optical counterparts can be identified for the Double, the Goldfish, and the Original TRG sources \citep[see also][]{miller03a2255}. The association of optical counterparts with the bright and filamentary emission due to the Trail and T-bone is not trivial. These sources are detected only with \lofar\ and trace some of the pixels with the steepest spectrum in the spectral index map (Fig.~\ref{fig:spix}). A possible connection between the E part of the Trail with a cluster member galaxy is noted. However, the surface brightness properties of the Trail are not similar to that of other tailed radio galaxies in the field (\eg\ the Goldfish and the Original TRG), in which there is a connection with the bright nuclear emission of the host galaxy (Fig.~\ref{fig:opt-radio}). This might suggest a different origin. The Trail extends towards W, where it mixes with the end of the Original TRG forming the emission named T-bone. The origin of these structures could be related to the interactions between member galaxies during the cluster merger. As a consequence of galaxy encounters, old relativistic electrons can be stripped and gain energy in the turbulent wakes produced behind galaxies moving through the ICM generating steep spectrum emission \citep[\eg][]{jaffe80, roland81coma}. Alternatively, they may trace fossil radio plasma revived by ICM motions \citep[\eg][]{ensslin01}, forming the so-called \textit{radio phoenixes} \citep[\eg][]{slee01, cohen11, vanweeren11nvss2, kale12relics, mandal20}; moreover, the possible connection between the T-bone and Original TRG could also indicate the presence of a Gently Re-Energized Tail \citep[GReET;][]{degasperin17gentle}. Regardless of their precise origin, the existence of ``radio ghosts'' in the ICM (\ie\ sources that are detected only at low frequencies, \citealt{ensslin99ghosts}) provides evidence of a population of seed electrons that could then be re-accelerated and spread into the cluster environment by ICM turbulence generated during the cluster merger event, as required by the leading models of radio halo formation \citep{brunetti01coma, petrosian01, brunetti07turbulence, donnert13, pinzke17}. The complex morphology and rapid surface brightness variations observed at the center of A2255 likely suggest that the radio plasma is dynamically interacting with velocity shears from turbulent motions and weak shocks in the ICM.
\\
\indent
The diffuse emission from the radio halo also displays distinctive surface brightness features. Elongated structures are detected in the form of the S spur and in the three straight filaments in the NW region (\cf\ Fig.~\ref{fig:a2255_multifreq}, central panel) that were already seen in higher frequency observations \citep{govoni05, pizzo11}. These filaments still have an unclear origin. They are embedded in the diffuse halo emission and show polarization levels of $\sim 20-40\%$ at 1.4 GHz \citep{govoni05}. A hypothesis advanced by \citet{pizzo11} from a deep rotation measure synthesis study is that the filaments may be related to radio relics located in the foreground of the cluster. The flatter spectral index values observed in the NW region (Fig.~\ref{fig:spix}) would be in agreement with this scenario as halos typically have steeper spectral indexes than relics. Therefore, despite the absence of obvious features in the X-ray image of this region, we find likely the possibility that we are observing a mixture of emission due to the superposition of the radio halo and other structures along the line of sight. Indeed, the presence of X-ray features (such as surface brightness jumps) can be hindered when radio relics are observed face-on. \\
\indent
Despite the remarkable features observed in the radio images, we found that the radio surface brightness is overall correlated with the X-ray surface brightness (Fig.~\ref{fig:a2255_ptp_sb}). Although the point-to-point analysis of radio halos has been investigated in about ten clusters \citep{govoni01comparison, feretti01, giacintucci05, brown11coma, shimwell14, rajpurohit18, hoang19a520, cova19, xie20}, results are still controversial. Firstly, the correlation can be tight \citep[\eg][]{govoni01comparison, feretti01} or very mild/absent \citep[\eg][]{shimwell14, cova19}. Secondly, the best-fit slopes vary over a broad range of values, from 0.25 \citep{hoang19a520} to 1.25 \citep{rajpurohit18}. In principle, the slope of the correlation could parameterize the particle acceleration mechanism \citep{dolag00, govoni01comparison, brunetti04nonthermal, pfrommer08gamma}. Different slopes (or the absence of a correlation) might also indicate that radio halos do not follow a universal scaling (see Section 5.3 in \citealt{shimwell14} for possible reasons), meaning that several mechanisms contribute to the acceleration mechanisms and transport of electrons. Moreover, we also do not exclude that different methods (\eg\ grid selection, linear regression procedure) used in the literature to derive the scalings may lead to dissimilar results (although the range of values appears too broad to entirely explain the slopes observed). In particular, we remark that setting a threshold in $\sigma$ on the radio measurements to calculate the linear regression introduces a selection bias that can affect the fitting of the data if not properly taken into account (Appendix~\ref{app:threshold}). \\
\indent
To summarize, we found a good relationship between the point-to-point radio and X-ray surface brightness of A2255. The correlation shows a large intrinsic scatter, which is reduced if grids with large cell sizes are used. The presence of this scatter demonstrates that the acceleration mechanisms are also sensitive to the details of local physical conditions (seeds and/or microphysics). We used \lira, which is able to treat the truncated distribution of data points generated by the $2\sigma$-cut adopted on the radio surface brightness, to fit the $I_R - I_X$ relation and obtained slopes that are consistent with linear scalings.

\subsection{Spectral properties}

The spectral index map presented in Fig.~\ref{fig:spix} shows a complex spectral index distribution spanning a wide range of values across the radio halo region. A clear contribution to the extreme flat and steep spectra are the extended radio galaxies embedded in the diffuse emission. The high surface brightness at low frequency of some of these features (\eg\ the Trail and the T-bone) in combination with their steep spectrum emission ($\alpha>2$) possibly indicate that they are pockets of radio plasma revived by ICM motions during the cluster merger (see also Section~\ref{sec:discussion_sb}). In our study, we focused on the diffuse emission. Besides the point-to-point analysis of the surface brightness, in Section~\ref{sec:spectra_ptp} we compared the spectral index of the diffuse emission and the fitting of X-ray spectra of the thermal ICM in the same cluster regions. \\
\indent
Firstly, we compared the spectral index versus the X-ray surface brightness, finding a mild positive trend (Fig.~\ref{fig:alpha_histo} and Appendix~\ref{app:threshold}). As the X-ray surface brightness declines towards the cluster outskirts, this result indicates a prevalence of steep spectrum emission in the the cluster center. Whilst old plasma emission from cluster central radio galaxies has been excluded in the analysis (Section~\ref{sec:spectra_ptp}), we can not rule out some level of residual contamination. Similar plots have been produced in the past \citep[\eg][]{shimwell14, hoang19a520} and in some cases azimuthally averaged spectral index profiles have been also reported \citep[\eg][]{orru07, vacca14, pearce17}. A possible, mild, radial steepening has been noticed for A2744 by \citet{pearce17}, whereas no clear trends are observed in the other cases\footnote{Earlier works also claimed the presence of spectral steepening in other systems \citep[\eg][]{giovannini93, feretti04spectral}; however, these studies were performed without matching the \uv-coverages of the instruments, making the results uncertain.}. Following the discussion in Appendix~\ref{app:threshold}, we remark that the threshold used to compute $\alpha$ combined with a large intrinsic scatter plays a role in the determination of the slope, in this case introducing possible observational biases at different frequencies (namely the different sensitivity of \lofar\ and \wsrt) that are difficult to take into account even using advanced regression algorithms. \\
\indent
Peculiar results on the halo in A2255 are also provided by the spectral index distribution, where a broad range of values appears statistically significant (Fig.~\ref{fig:alpha_histo}, right panels and Appendix~\ref{app:threshold}). To the best of our knowledge, possible complexity in the spectral index distribution has been quantitatively reported only by \citet{pearce17}, which could attribute to measurement errors only $\sim50\%$ of the dispersion of the spectral index distribution of the halo in A2744. The distribution of spectral index obtained is projected along the line of sight and has a dispersion $\sigma_{\alpha,\,proj} \sim 0.3$. If we assume a scenario where $\alpha$ varies in the volume of the halo, the observed distribution provides only a lower limit to the intrinsic distribution of spectral index values. More quantitatively, following \citet{hoang19a520}, if we assume that $\alpha$ changes stochastically on a spatial scale $L$, the intrinsic scatter is $\sigma_{\alpha,\,intr} \sim \sigma_{\alpha,\,proj} \times \sqrt{N_L}$, where $N_L$ is the number of independent cells of size $L$ that are intercepted along the line of sight. In Appendix~\ref{app:threshold}, we have shown that the scatter does not change dramatically with the cell size, implying (in the above scenario) that we are in the condition where the scale $L$ is similar or larger than cell size ($\gtrsim 50-70$ kpc), implying an intrinsic scatter in the range $\sigma_{\alpha,\,intr} \sim 0.6-0.9$. 
This is a very large scatter that would indicate a very inhomogeneous interplay between acceleration rate and seed particles across the halo volume. Conversely, a remarkable low scatter is observed for the radio halos in the Toothbrush cluster \citep{vanweeren16toothbrush} and in A520 \citep{hoang19a520}. These differences may suggest that the turbulent energy is not homogeneously dissipated in the halo volume and that different time-scales and dynamical states in the merger can lead to diverse properties of the turbulence in the ICM. As a final remark, we note that we also did not find a significant dependency of the dispersion of $I_R$ with the cell size, indicating a similar scale of spatial variation of electrons per unit volume. \\ 
\indent
As last step of our analysis, we searched for a connection between radio and X-ray spectral information. In this case, the analysis was done in a smaller number of cluster regions (with larger sizes) to overcome the problem of the different SNR of the three datasets. A comparison between spatially resolved spectral index and temperature values has been firstly reported in \citet{orru07}, who found that the regions with highest gas temperature trace the flatter spectrum emission of the radio halo in A2744. Recently, this result was called into question by \citet{pearce17}, who used deeper radio and X-ray data for the same cluster and found no strong evidence for the existence of a correlation between the two quantities (see also \citealt{shimwell14} for the case of the Bullet cluster). We find a negative trend between temperature and spectral index for A2255 and, for the first time, we also searched for a possible relation between $\alpha$ and pseudo-pressure/pseudo-entropy (Fig.~\ref{fig:alpha_thermo}). We do not find strong evidence for a tight a $P-\alpha$ correlation but highlight an anti-correlation between spectral index and pseudo-entropy. Both higher temperature and high pseudo-entropy are expected to trace regions where gas has been heated and mixed by shocks and turbulent motions \citep[\eg][]{gaspari14, shi20}. Therefore, one may suppose that the dynamics of the gas and the turbulence (more precisely, the turbulent energy flux) that is damped into particles correlates with entropy making re-acceleration more efficient (and thus producing radio emission with flatter spectrum) in higher entropy regions. On the other hand, we can not exclude that this trend comes from the combination of the increasing entropy and temperature profile towards cluster outskirts and excess of steep spectrum emission in the cluster center (Fig.~\ref{fig:alpha_histo}). The investigation of these possible correlations and their physical implications require detailed studies that are beyond the scope of this paper.

\section{Conclusions}\label{sec:conclusions}

We have presented \lofar\ observations at $120-168$ MHz of the nearby merging galaxy cluster A2255. The images reported in the paper are among the deepest ever obtained on a galaxy cluster at low frequencies; remarkably, these come from a large-area sky survey, that is \lotss. The picture emerging from our work is that of one of the most intricate diffuse radio sources known to date. In the central $\sim10$ Mpc$^2$ region, radio emission is found spanning a wide range of physical scales, and is mainly contributed by diffuse emission from the ICM and extended emission from tailed radio galaxies. Among the new features discovered by \lofar, we highlight the presence of filamentary and distorted structures on different scales with very steep spectrum ($\alpha>2$) embedded in the radio halo. These observations are suggestive of a complex gas dynamics including shear and turbulent motions and (possibly) weak shocks in the ICM. \\
\indent
We used archival \wsrt\ observations at 1.2 GHz and \chandra\ data to investigate the spectral properties of the cluster and to study the connection between thermal and non-thermal components in the ICM. \\
\indent
The point-to-point comparison between the X-ray and radio surface brightness of the ICM indicates that the two are correlated. Our results are consistent with linear scalings and have been obtained by taking into account the presence of the $2\sigma$-cut used to evaluate the radio surface brightness. We remark that the slope of the correlation in case of large intrinsic scatter can be biased if this selection effect is not treated by the fitting algorithm. \\
\indent
We found that the radio halo displays a broad range of spectral index values, possibly indicating a mix of populations of seed electrons with different properties injected by cluster radio galaxies and distributed and re-accelerated in the ICM by merger-induced turbulent motions. A mild trend observed between the X-ray surface brightness and radio spectral index suggests a prevalence of steep spectrum emission in the innermost cluster region. We also compared the spectral index with the thermodynamical quantities of the ICM evaluated in the same cluster regions. Whilst we could not firmly claim strong relations between spectral index and temperature/pseudo-pressure, we found possible evidence for an anti-correlation between $\alpha$ and pseudo-entropy. If confirmed, this could indicate a connection between spectrum of the radio emission and turbulent energy flux or heating and mixing of the gas in the ICM. We believe that the study of these correlations (or the lack of) deserves further investigation in the future. \\
\indent
In this paper we focused on the central region of A2255 using three pointings coming from \lotss. The analysis of other features in the system, such as the extended emission at large distance from the cluster center, will be performed in forthcoming works which will exploit deep (75~hr), pointed \lofar\ observations of the cluster. 

\acknowledgments

We thank the anonymous referee for helpful comments and A. Ignesti and M. Sereno for useful discussions during the preparation of this paper. ABot and RJvW acknowledge support from the VIDI research programme with project number 639.042.729, which is financed by the Netherlands Organisation for Scientific Research (NWO). ABon acknowledges support from ERC-Stg DRANOEL 714245 \& MIUR FARE SMS. GDG acknowledges support from the ERC Advanced Investigator programme NewClusters 321271. AD acknowledges support by the BMBF Verbundforschung under the grant 05A17STA. HJAR acknowledges support from the ERC Advanced Investigator programme NewClusters 321271. LOFAR \citep{vanhaarlem13} is the LOw Frequency ARray designed and constructed by ASTRON. It has observing, data processing, and data storage facilities in several countries, which are owned by various parties (each with their own funding sources), and are collectively operated by the ILT foundation under a joint scientific policy. The ILT resources have benefitted from the following recent major funding sources: CNRS-INSU, Observatoire de Paris and Universit\'{e} d'Orl\'{e}ans, France; BMBF, MIWF-NRW, MPG, Germany; Science Foundation Ireland (SFI), Department of Business, Enterprise and Innovation (DBEI), Ireland; NWO, The Netherlands; The Science and Technology Facilities Council, UK; Ministry of Science and Higher Education, Poland; Istituto Nazionale di Astrofisica (INAF), Italy. This research made use of the Dutch national e-infrastructure with support of the SURF Cooperative (e-infra 180169) and the LOFAR e-infra group, and of the LOFAR IT computing infrastructure supported and operated by INAF, and by the Physics Dept.~of Turin University (under the agreement with Consorzio Interuniversitario per la Fisica Spaziale) at the C3S Supercomputing Centre, Italy. The J\"{u}lich LOFAR Long Term Archive and the German LOFAR network are both coordinated and operated by the J\"{u}lich Supercomputing Centre (JSC), and computing resources on the supercomputer JUWELS at JSC were provided by the Gauss Centre for Supercomputing e.V. (grant CHTB00) through the John von Neumann Institute for Computing (NIC). This research made use of the University of Hertfordshire high-performance computing facility and the LOFAR-UK computing facility located at the University of Hertfordshire and supported by STFC [ST/P000096/1]. The scientific results reported in this article are based in part on data obtained from the \chandra\ Data Archive. SRON Netherlands Institute for Space Research is supported financially by the Netherlands Organisation for Scientific Research (NWO). This research made use of APLpy, an open-source plotting package for Python \citep{robitaille12}. \facilities{\lofar, \wsrt, \chandra}


\appendix

\section{Grids for point-to-point analysis}\label{app:grids}

\begin{figure*}[h]
 \centering
 \includegraphics[width=.33\hsize,trim={0.9cm 0cm 0.9cm 0cm},clip]{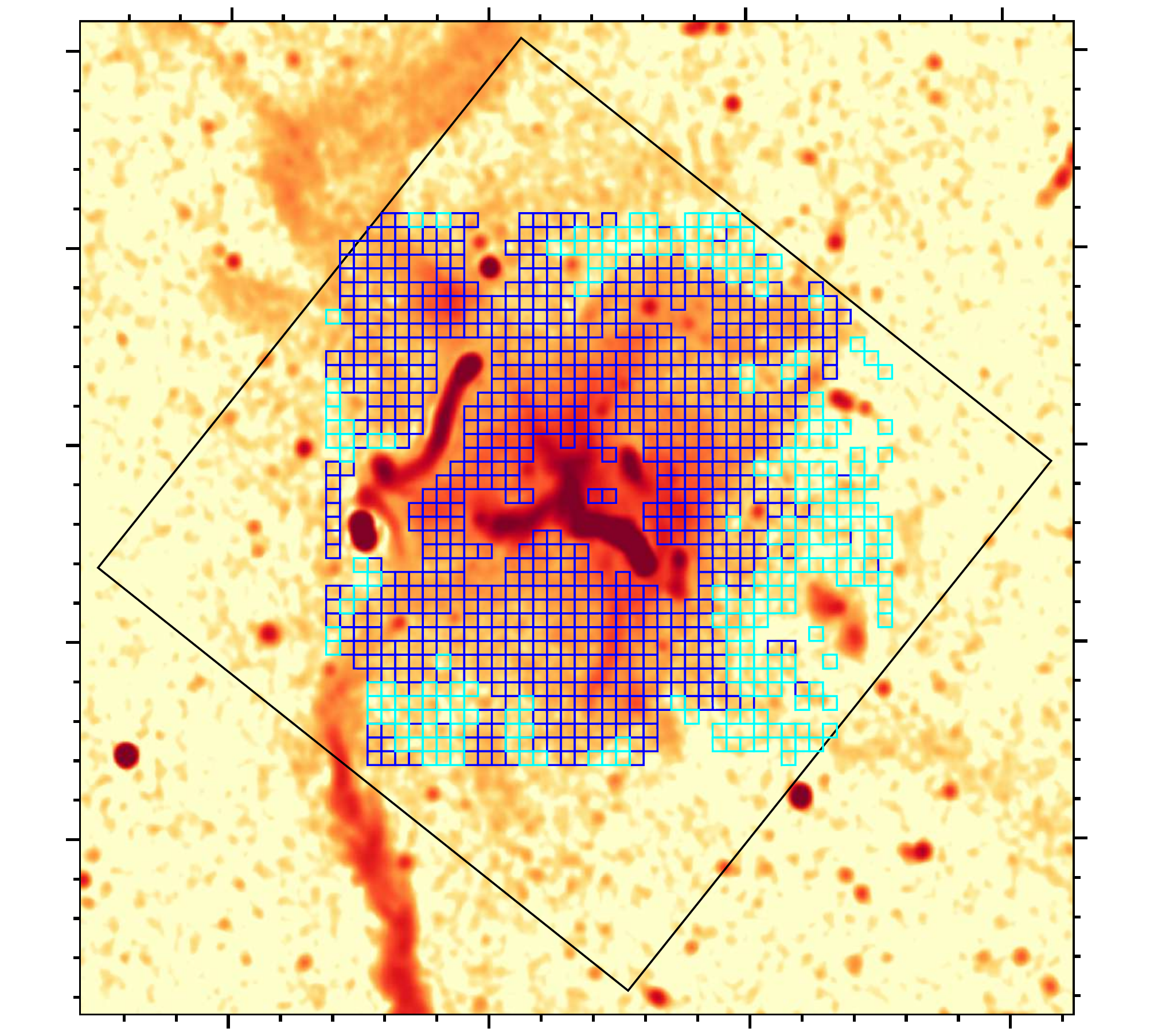}
 \includegraphics[width=.33\hsize,trim={0.9cm 0cm 0.9cm 0cm},clip]{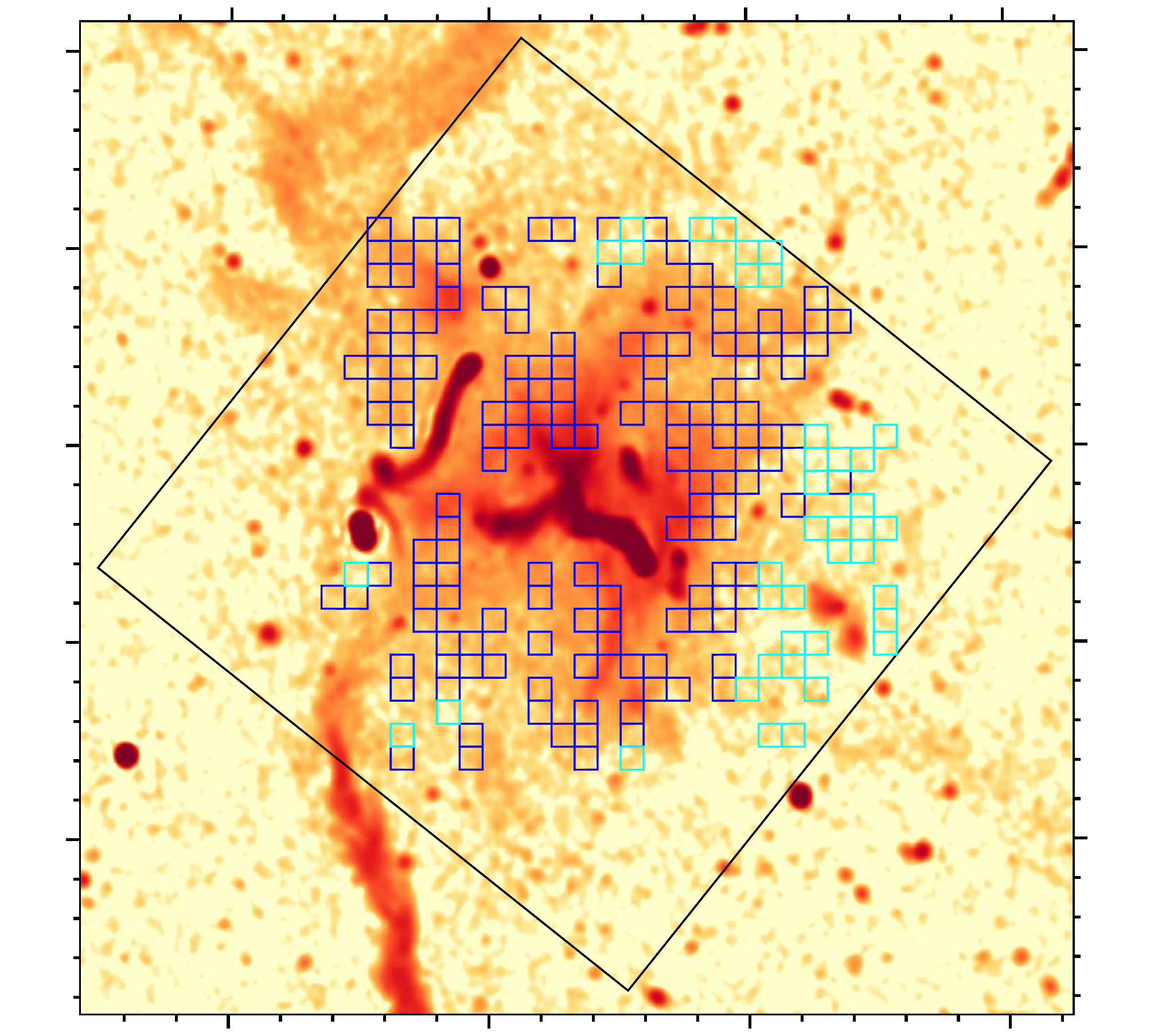}
 \includegraphics[width=.33\hsize,trim={0.9cm 0cm 0.9cm 0cm},clip]{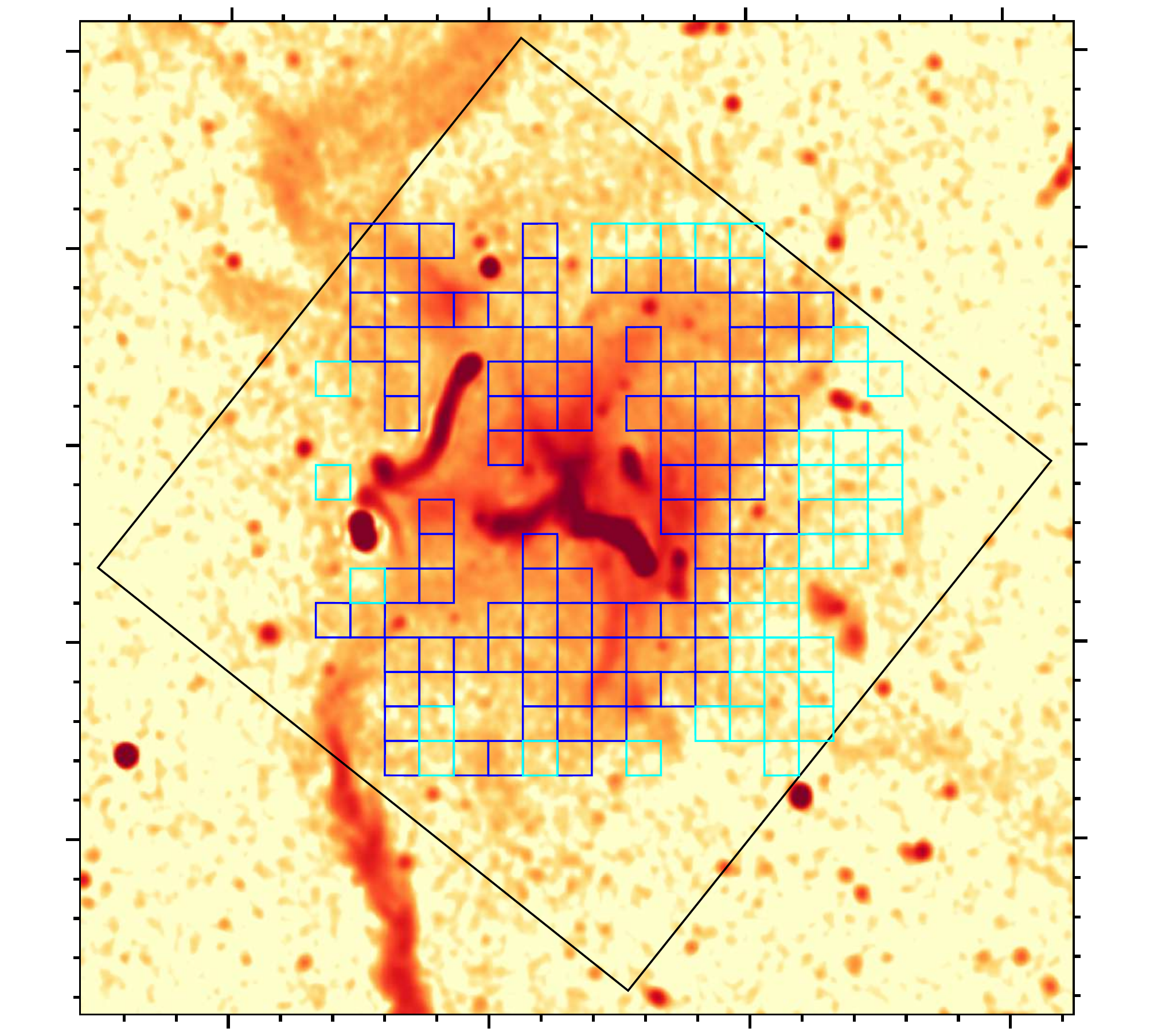}
 \caption{Grids used for the point-to-point analysis overlaid on the \lofar\ images at $15\arcsec \times 14\arcsec$ resolution. From \textit{left} to \textit{right}, the displayed cells are those used to compute the values in Fig.~\ref{fig:a2255_ptp_sb} ($21\arcsec \times 21\arcsec$, $35\arcsec \times 35\arcsec$, and $52.5\arcsec \times 52.5\arcsec$). Blue and cyan cells mark regions where the radio emission is above or below $2\sigma$, respectively. The black region denotes the \acis-I FoV.}
 \label{fig:grids}
\end{figure*}

The three grids with different cell sizes used to evaluate the radio and X-ray surface brightness for the point-to-point analysis are reported in Fig.~\ref{fig:grids}.

\section{Impact of the threshold on the radio analysis}\label{app:threshold}

\begin{figure*}[h]
 \centering
 \includegraphics[width=.29\hsize,trim={1.0cm 0.2cm 1.0cm 0.2cm},clip]{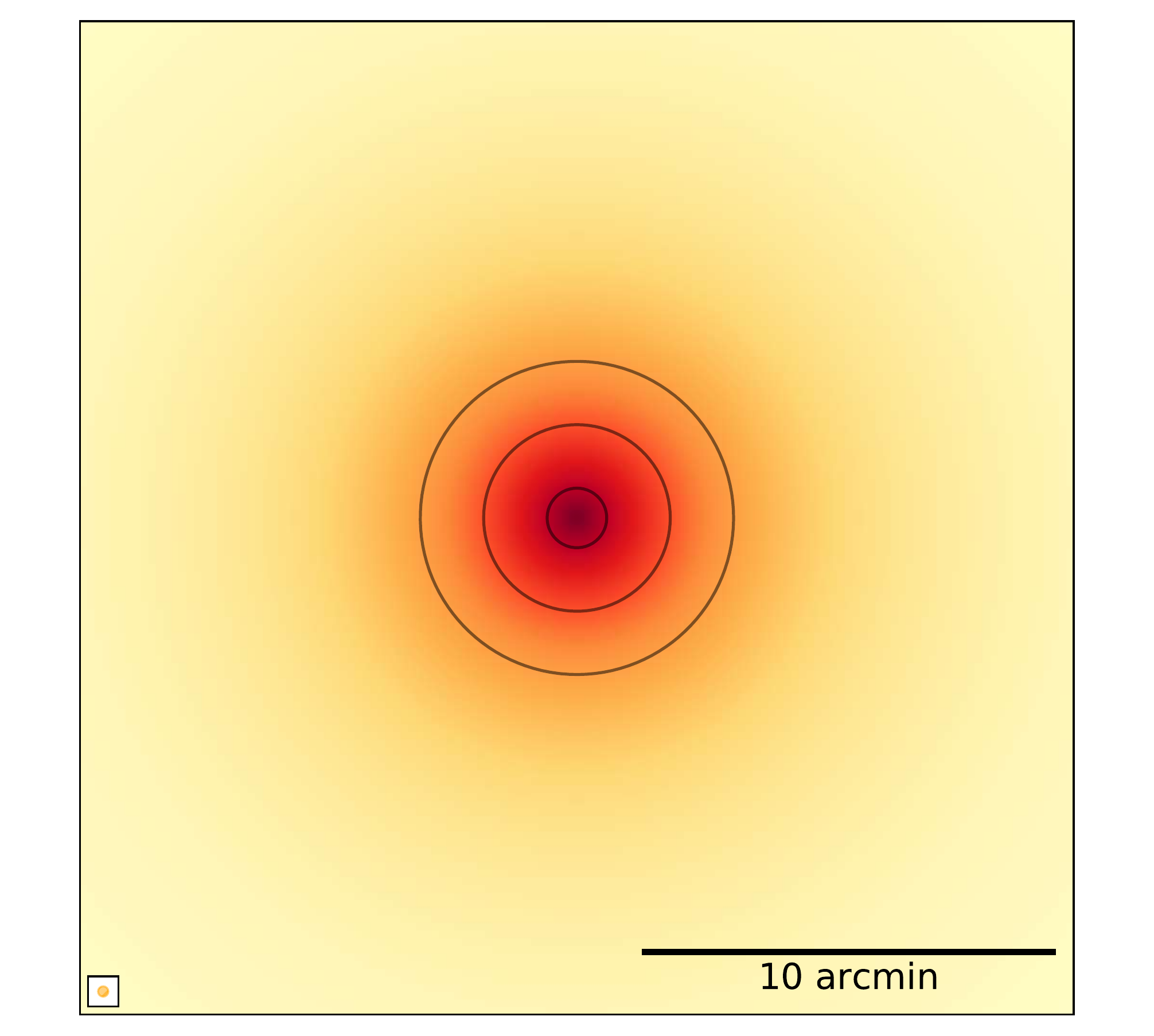}
 \includegraphics[width=.29\hsize,trim={1.0cm 0.2cm 1.0cm 0.2cm},clip]{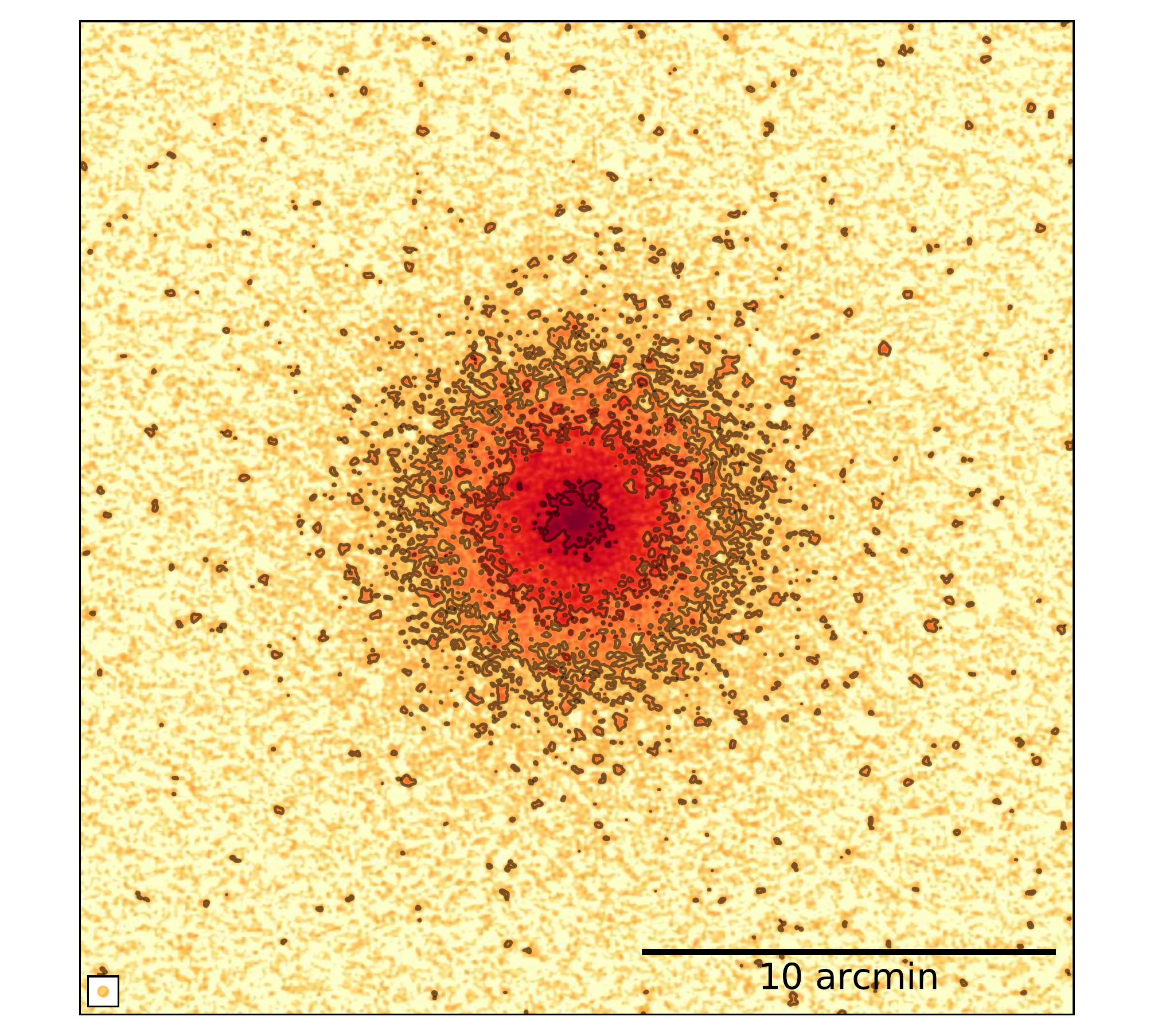}
 \includegraphics[width=.37\hsize,trim={6.5cm 9.0cm 6.5cm 1.1cm},clip]{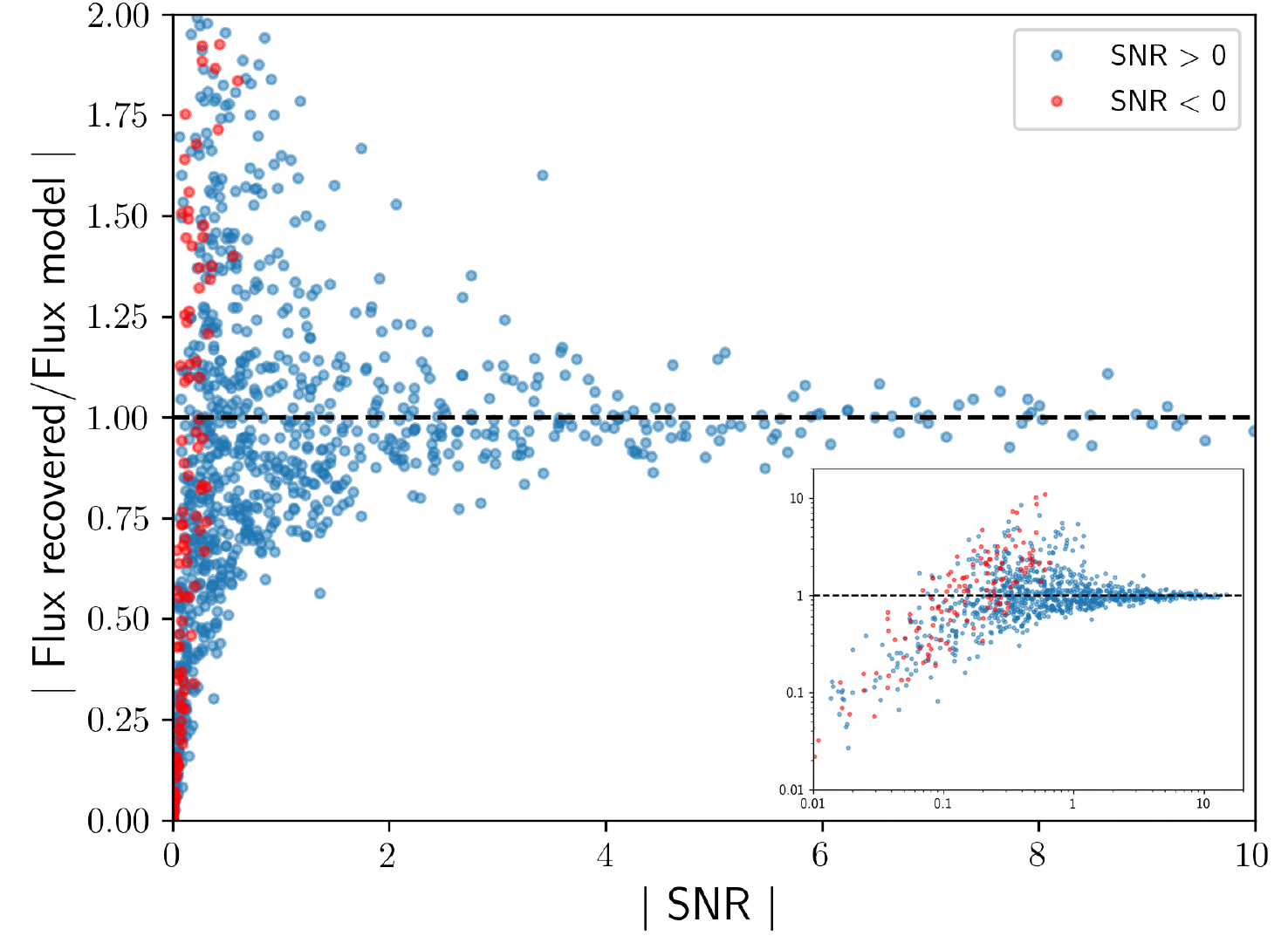}
 \caption{Exponential radio halo model injected in the visibilities (\textit{left}) and corresponding restored image (\textit{center}) at the same resolution of $15\arcsec \times 15\arcsec$. The colorscale in the two images is matched and contours/circles are drawn at levels of $120 \times [3, 6, 12]$ \mujyb. The absolute value of the ratio between recovered flux and injected flux as a function of the absolute value of the SNR (\textit{right}) has been computed on a point-to-point basis from the images shown in the other two panels. The inset shows a zoom-out of the panel plotted in log-scale.}
 \label{fig:injection}
\end{figure*}

\begin{figure}[t]
 \centering
 \includegraphics[width=0.5\hsize,trim={0cm 0cm 0cm 0cm},clip]{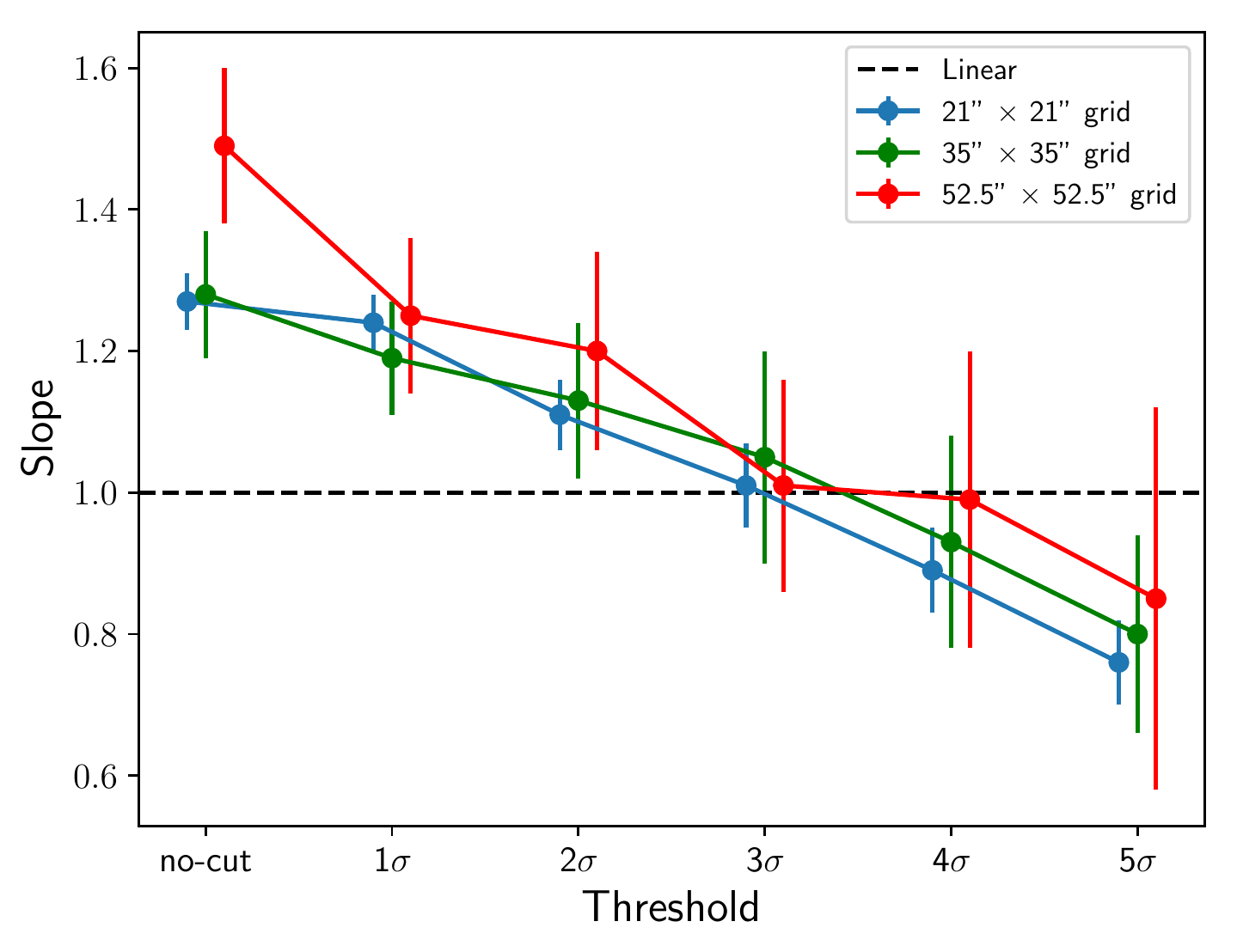}
 \caption{Slope of the $I_R-I_X$ relation obtained with the orthogonal-BCES method as function of the radio threshold used to fit the data for the three different grids. Points are slightly offset on the \textit{x}-axis for visualization purposes.}
 \label{fig:slopes}
\end{figure}

\begin{figure*}[ht]
 \centering
 \includegraphics[width=.26\hsize]{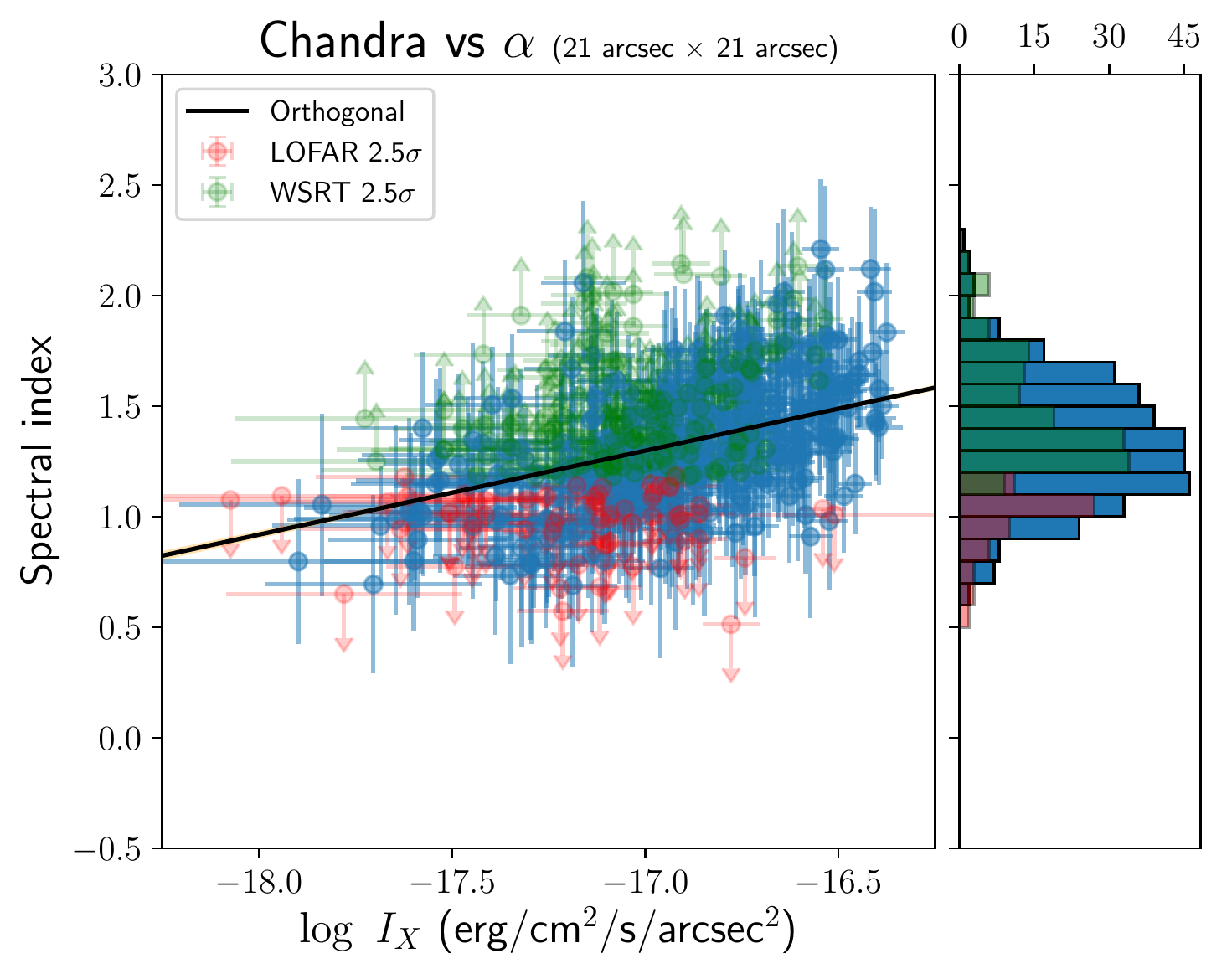}
 \includegraphics[width=.26\hsize]{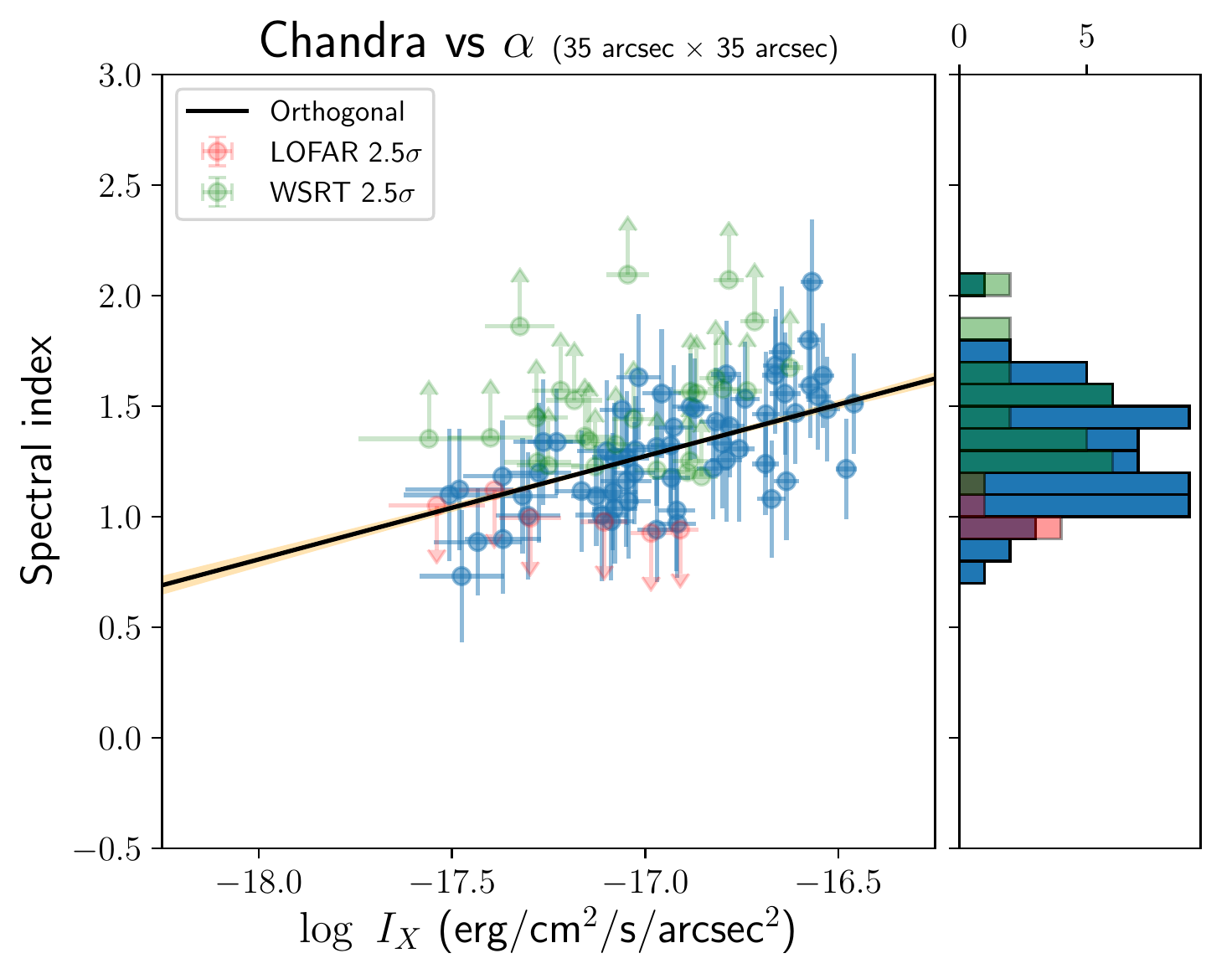}
 \includegraphics[width=.26\hsize]{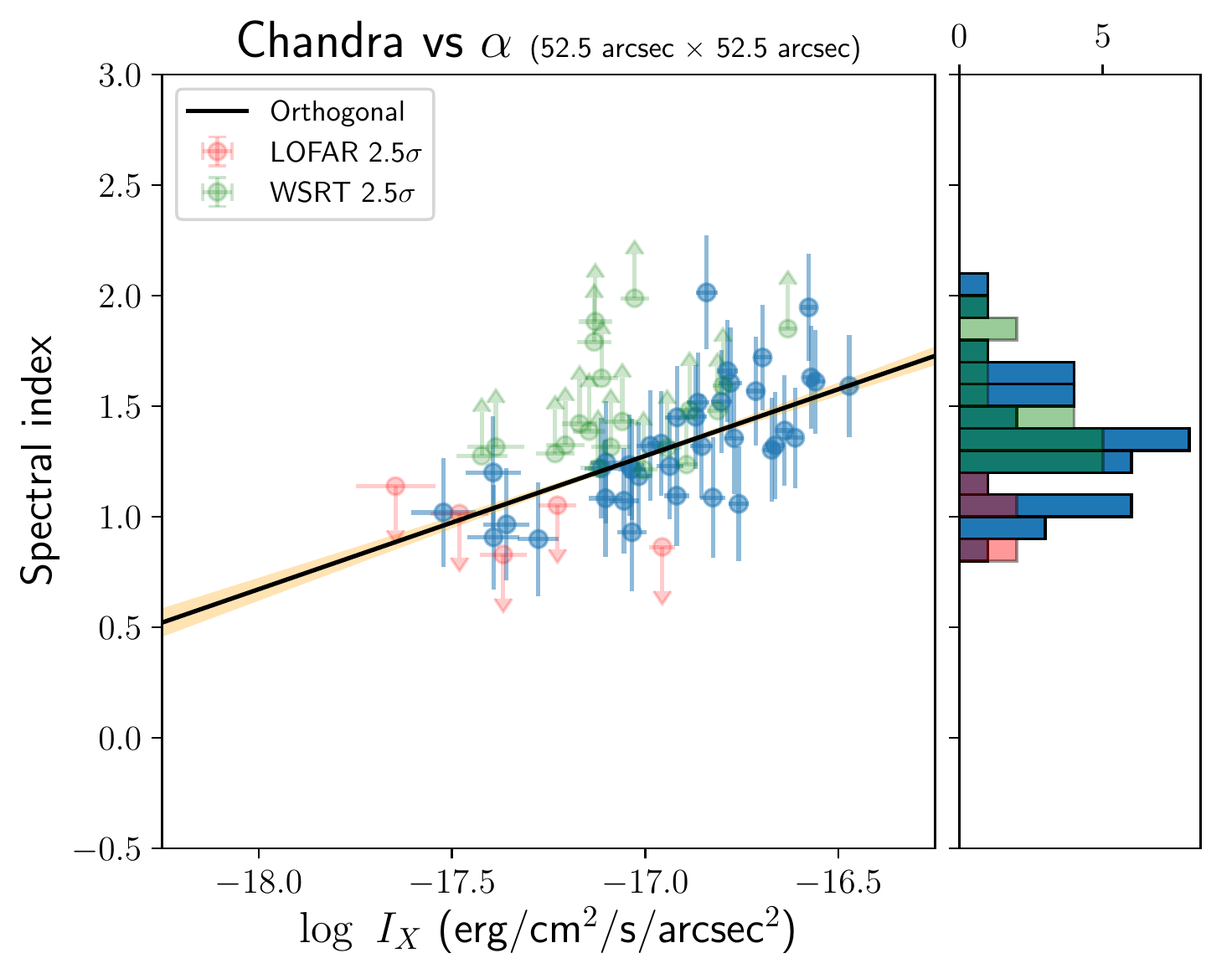} \\
 \includegraphics[width=.26\hsize]{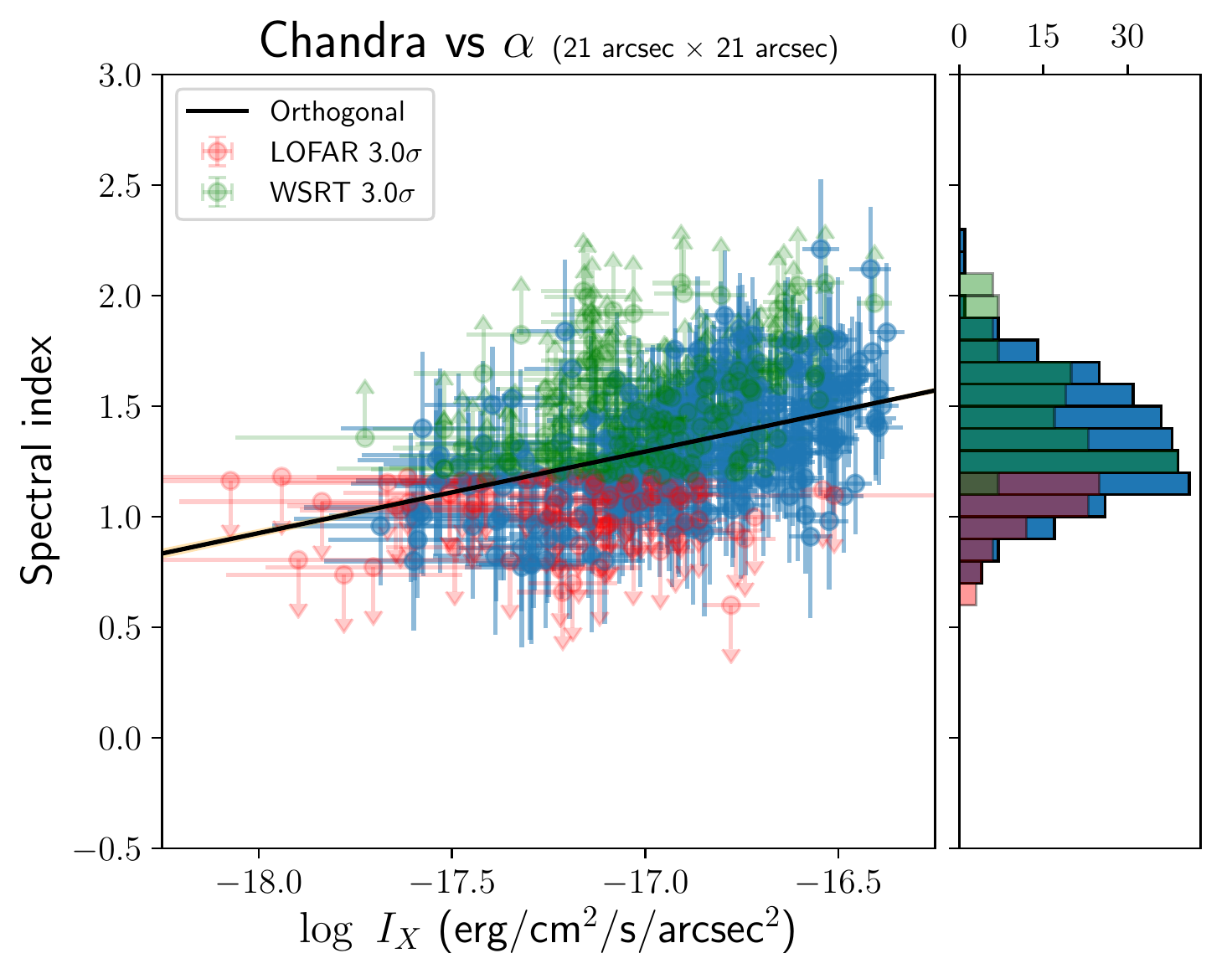}
 \includegraphics[width=.26\hsize]{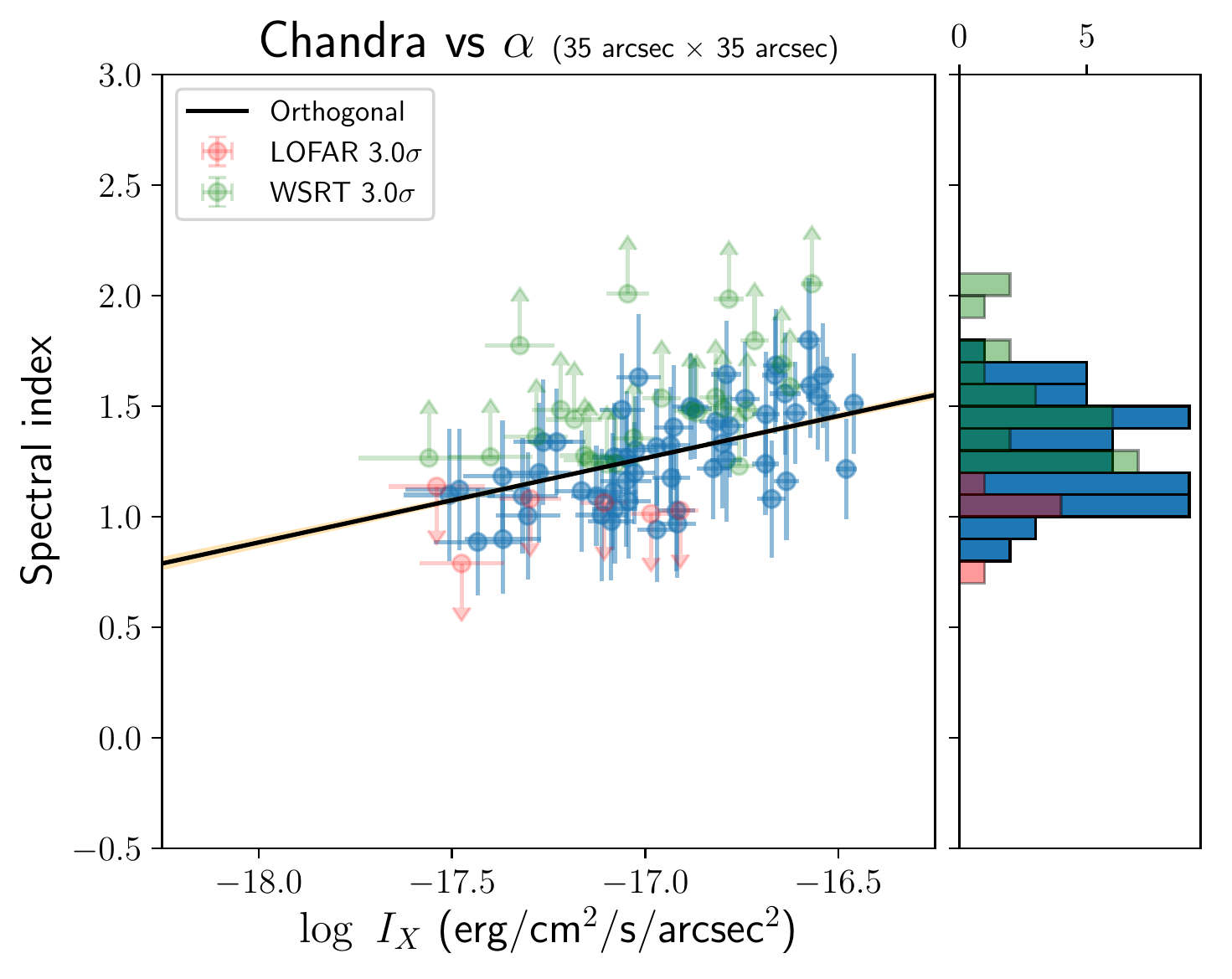}
 \includegraphics[width=.26\hsize]{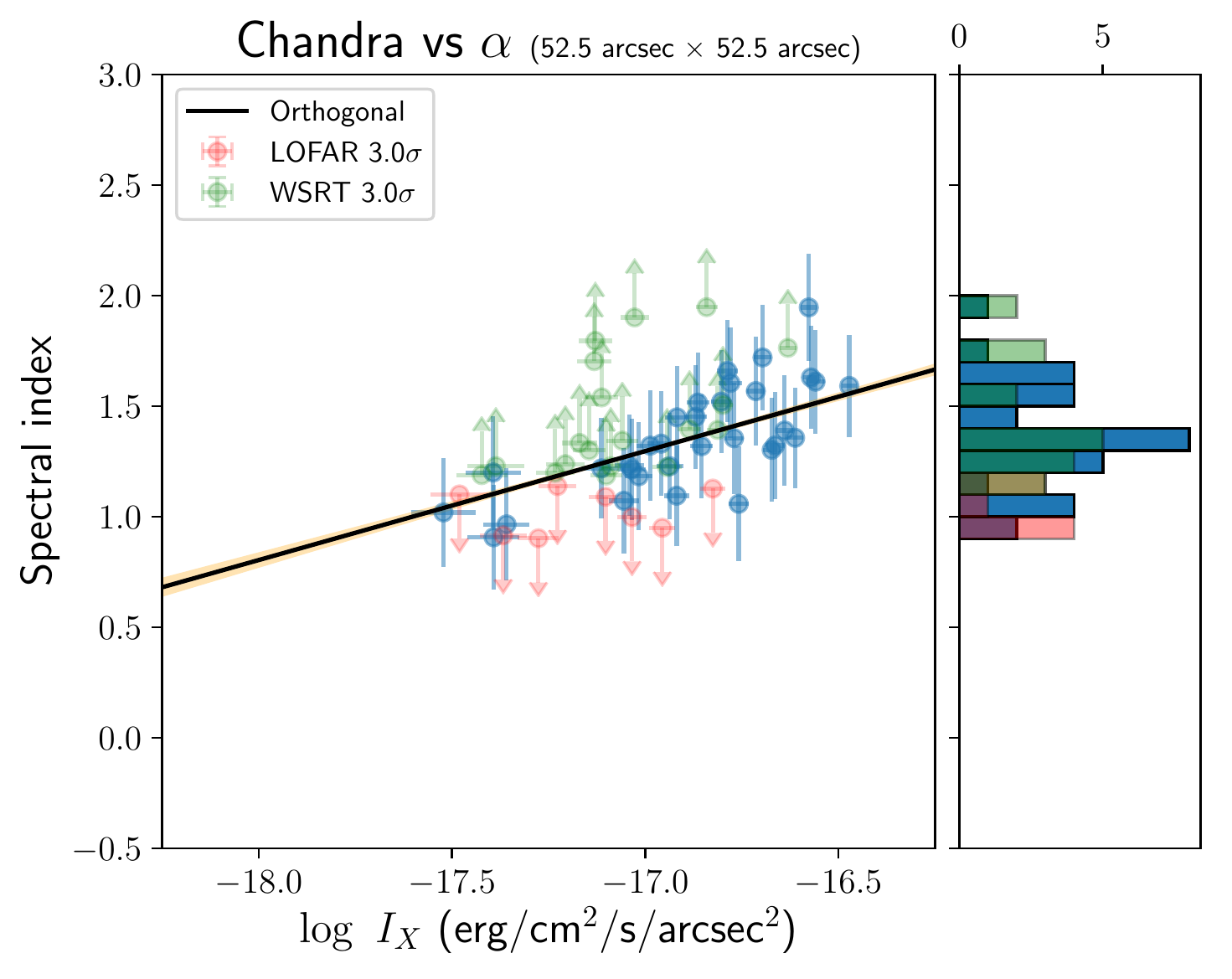}
 \caption{Same as Fig.~\ref{fig:alpha_histo}, but for thresholds of $2.5\sigma$ (\textit{top}) and $3\sigma$ (\textit{bottom}) and for different grids (from \textit{left} to \textit{right}).}
 \label{fig:alpha_histo_threshold}
\end{figure*}

Artifacts in interferometric radio images arising in process of image reconstruction can bias the flux density estimates especially for extended sources with low-SNR. We quantify this effect by using our \lofar\ observations of A2255. First, we created a model of radio halo with an exponential surface brightness profile with a similar extension to that observed in A2255 (Fig.~\ref{fig:injection}, left panel). Second, we Fourier-transformed the image of the mock halo and added it in the \lofar\ visibilities consisting of Gaussian random noise. This radio halo model was restored using the same \wsclean\ parameters adopted in the \lofar\ imaging, leading to the recovered image shown in Fig.~\ref{fig:injection} (central panel). We used these two images to compare on a point-to-point basis the flux density recovered in the restored image with that in the model image as a function of the SNR (of the restored image). As negatives flux densities are possible in radio images due to artifacts and random thermal noise fluctuations, we plotted the absolute values of the ratio between the two flux densities and SNR (basically, $\rm{SNR}<0$ when the the observed flux density is negative). Results are shown in Fig.~\ref{fig:injection} (right panel) and show that at high-SNR the emission of the mock radio halo is well recovered (\ie\ the ratio between the flux densities is $\sim1$), while a large scatter is evident at low-significance ($\lesssim1\sigma$). The same trend is observed by varying by a factor of 2 the $e$-folding radius of the injected halo. This demonstrates that the flux densities observed in radio images at low-SNR are not reliable, and that only flux densities above a certain threshold should be considered in the analysis. In the case of real observations, where radio halo morphologies are more complex than an exponential profile and the noise is possibly not Gaussian, our results suggest that a threshold of $\gtrsim2\sigma$ should be applied when measuring radio flux densities. \\
\indent
Although the general approach that has been used in the literature to compare the radio and X-ray surface brightness of clusters hosting radio halos was to adopt a threshold on the radio surface brightness, we note that this choice can introduce a bias in the determination of the $I_R - I_X$ slope in case of large intrinsic scatter. We prove this by performing linear on Eq.~\ref{eq:ptp_sb} using the bivariate correlated errors and intrinsic scatter (BCES) orthogonal method \citep{akritas96} and by considering different values of the threshold. Results are summarized in Fig.~\ref{fig:slopes}. The trend obtained between $I_R$ and $I_X$ is overall consistent between the different grids and demonstrates that applying higher thresholds on $I_R$ leads to flatter slopes. The reason of this is that close to the detection threshold the selection picks up only the up-scattered values, which leads to an overestimate of the mean value of $I_R$ at low $I_X$ and an underestimate of the slope. This motivates the choice of a Bayesian regression method, such as \lira\ \citep{sereno16}, which is able to take into account the effect of this Malmquist bias. \\
\indent
We now investigate the impact of the threshold adopted on the computation of the spectral index distribution and on the correlation $\alpha-I_X$ presented in Section~\ref{sec:spectra_ptp}. In Fig.~\ref{fig:alpha_histo_threshold} we reproduce the plots shown in Fig.~\ref{fig:alpha_histo} by adopting thresholds of 2.5$\sigma$ and 3$\sigma$. Results are summarized in Tab.~\ref{tab:threshold}. First, we note that there is only a small variation of the mean spectral index value $\overline{\alpha}$ and its standard deviation $\sigma_\alpha$ with the thresholds and grids used. Second, we notice that the spectral index is always positively correlated with the X-ray surface brightness, and that this trend would not change even assuming the most conservative values for the upper and lower limits. In this case, we warn the reader that the reported slopes should be taken with caution as they were derived without taking into account the limits on the spectral indexes.

\begin{table}[h]
 \centering
 \caption{Spectral index mean value ($\overline{\alpha}$) with its standard deviation ($\sigma_\alpha$), slope ($A$) from the BCES-orthogonal fitting to Eq.~\ref{eq:ptp_alpha}, and Spearman ($r_s$) and Pearson ($r_p$) correlation coefficients relative to the plots shown in Fig.~\ref{fig:alpha_histo} and Fig.~\ref{fig:alpha_histo_threshold}.}
 \label{tab:threshold}
  \begin{tabular}{lccccc} 
  \hline
  \hline
  Threshold & $\overline{\alpha}$ & $\sigma_\alpha$ & $A$ & $r_s$ & $r_p$ \\ 
  \hline
  & \multicolumn{5}{c}{$21\arcsec\times21\arcsec$} \\
  \hline
  $2.0\sigma$ & 1.34 & 0.30 & $0.34\pm0.03$ & 0.50 & 0.48 \\
  $2.5\sigma$ & 1.33 & 0.28 & $0.38\pm0.03$ & 0.55 & 0.55 \\ 
  $3.0\sigma$ & 1.34 & 0.26 & $0.37\pm0.04$ & 0.55 & 0.55 \\ 
  \hline
  & \multicolumn{5}{c}{$35\arcsec\times35\arcsec$} \\
  \hline
  $2.0\sigma$ & 1.30 & 0.25 & $0.41\pm0.08$ & 0.63 & 0.61 \\ 
  $2.5\sigma$ & 1.30 & 0.26 & $0.47\pm0.08$ & 0.65 & 0.66 \\ 
  $3.0\sigma$ & 1.29 & 0.23 & $0.38\pm0.06$ & 0.63 & 0.63 \\ 
  \hline
  & \multicolumn{5}{c}{$52.5\arcsec\times52.5\arcsec$} \\
  \hline
  $2.0\sigma$ & 1.32 & 0.27 & $0.51\pm0.10$ & 0.69 & 0.66 \\ 
  $2.5\sigma$ & 1.33 & 0.27 & $0.60\pm0.12$ & 0.73 & 0.69 \\
  $3.0\sigma$ & 1.36 & 0.24 & $0.49\pm0.09$ & 0.74 & 0.74 \\ 
 \hline
  \end{tabular}
\end{table}


\bibliographystyle{aasjournal}
\bibliography{library.bib}



\end{document}